\documentclass[review]{elsarticle}

\usepackage[utf8]{inputenc}
\usepackage[T1]{fontenc}
\usepackage{tgtermes}
\usepackage{lineno,hyperref,pifont,natbib,geometry,fleqn,graphicx,makeidx,fancyhdr,multirow,verbatim,amsmath,amsthm,amssymb,float,array,subfloat,epstopdf,subfig}
\modulolinenumbers[5]

\journal{Journal of Instrumentation}

\begin{document}

\begin{frontmatter}

% \title{Elsevier \LaTeX\ template\tnoteref{mytitlenote}}
% \title{Characterization of Thin p-on-p Radiation Detectors with Active Edges} % -\\ RD50 Collaboration Status Report} % \tnoteref{t1}}
% \title{Signal formation in segmented Si planar detectors: TCAD simulated effect of SiO$_2$ passivation layer}
\title{Simulation study of signal formation in position sensitive planar p-on-n silicon detectors after short range charge injection{\large *}}\corref{cor1}
% \title{This is a specimen title\tnoteref{t1}}
 % \tnotetext[t1]{On behalf of the RD50 Collaboration, A complete author list can be found at: http://rd50.web.cern.ch/rd50.}
% \tnotetext[mytitlenote]{Fully documented templates are available in the elsarticle package on \href{http://www.ctan.org/tex-archive/macros/latex/contrib/elsarticle}{CTAN}.}

% \address{Radarweg 29, Amsterdam}
% \fntext[myfootnote]{Since 1880.}
 % \author{Timo Peltola\corref{mycorrespondingauthor}\fnref{myfootnote2}}
 % \address{Helsinki Institute of Physics, P.O.Box 64 (Gustaf H\"{a}llstr\"{o}min katu 2) FI-00014 University of Helsinki, FINLAND}

% \author[rvt]{Timo Peltola\corref{cor1}\fnref{fn1}}
% \author[rvt]{Timo Peltola\corref{cor1}}
\author[a,b]{T. Peltola\corref{cor2}}
\ead{timo.peltola@helsinki.fi}
\author[c]{V. Eremin} %\corref{cor2}}
% \ead{Xiaopeng.Wu@vtt.fi} 
\author[c]{E. Verbitskaya}
% \author[d]{C. Granja}
% \author[d]{J. Jakubek}
% \author[d]{M. Jakubek}
\author[d,b]{J. H\"{a}rk\"{o}nen}
% \author[a]{A. G\"{a}dda}
% \author[c]{A. Junkes} \author[d]{P. Kostamo} \author[e]{H. Lipsanen} \author[e]{M. Mattila} \author[d]{S. Nenonen} \author[a]{T. Riekkinen} 

\address[a]{Texas Tech University, Department of Physics and Astronomy, Lubbock, TX, 79409, USA}
\address[b]{Helsinki Institute of Physics, P.O. Box 64 (Gustaf H\"{a}llstr\"{o}min katu 2) FI-00014 University of Helsinki, Finland}
% \address[c]{Ioffe Physico-Technical Institute or Russian Academy of Sciences, St. Petersburg 194021, Russia}
\address[c]{Ioffe Institute, St. Petersburg 194021, Russian Federation}
\address[d]{Ru\dj{er} Bo\v{s}kovi\'{c} Institute, Zagreb, 10000, Croatia}
% \address[d]{Ruder Bo\v{s}kovi\'{c} Institute, Zagreb, 10000, Croatia}
% \address[b]{VTT, Microsystems and Nanoelectronics, Tietotie 3, Espoo, P.O. Box 1000, FI-02044 VTT, Finland}
% \address[c]{Advacam Oy, Tietotie 3, Espoo, FI-02150, Finland} 
% \address[d]{Institute of Experimental and Applied Physics, Czech Technical University in Prague (IEAP-CTU), Horsk\'{a} 3a/22, CZ 12800 Prague 2, Czech Republic} 

% \address[e]{Aalto University, Department of Micro and Nanosciences, Tietotie 3, Espoo, P.O. Box 13500, FI-02015, Finland}
% \author{\small On behalf of the RD50 Collaboration\fnref{myfootnote}}
% \fntext[myfootnote]{Speaker.}
% \fntext[myfootnote]{A complete author list can be found at: http://rd50.web.cern.ch/rd50.}
% \ead[url]{http://rd50.web.cern.ch/rd50.}

% \author[mymainaddress,mysecondaryaddress]{Elsevier Inc}

% \author[mysecondaryaddress]{Global Customer Service\corref{mycorrespondingauthor}}
% \cortext[mycorrespondingauthor]{Speaker.}
\cortext[cor1]{Work performed in the framework of the CERN-RD50 collaboration.}
\cortext[cor2]{Corresponding author}
% \cortext[cor2]{Corresponding author}
% \fntext[fn1]{On behalf of the RD50 collaboration, http://rd50.web.cern.ch/rd50.}

 % \address[rvt]{Helsinki Institute of Physics, P.O.Box 64 (Gustaf H\"{a}llstr\"{o}min katu 2) FI-00014 University of Helsinki, Finland}
 % \ead{timo.peltola@helsinki.fi}

% \cortext[mycorrespondingauthor2]{On behalf of the RD50 collaboration.}
% \ead{http://rd50.web.cern.ch/rd50.}

% \address[mymainaddress]{1600 John F Kennedy Boulevard, Philadelphia}
% \address[mysecondaryaddress]{360 Park Avenue South, New York}

\begin{abstract}
% This template helps you to create a properly formatted \LaTeX\ manuscript.
\label{Abstract}
% The transient current technique has been used to investigate signal formation in unirradiated silicon microstrip detectors, which are similar in geometry to those developed for the ATLAS experiment at LHC. % Nanosecond pulsed infrared and red lasers were used to induce the signals under study. Two peculiarities in the detector performance were observed: % an unexpectedly slow rise to the signal induced in a given strip when signals are injected opposite to the strip, and a long duration of the induced signal in comparison with the calculated drift time of charge carriers through the detector thickness with a significant fraction of the charge being induced after charge carrier arrival. These major effects and details of the detector response for different positions of charge injection are discussed in the context of Ramo's theorem % and compared with predictions arising from the more commonly studied phenomenon of signal formation in planar pad detectors. (Eremin)\\ \\
%%%%%%%%%%%%%%%%%%%%%%%%%%%%%%%%%%%%%%%%%%%%%%%%
Segmented silicon detectors 
(micropixel and microstrip) are the main type of detectors used in the inner
trackers of Large Hadron Collider (LHC) experiments at CERN. Due to the
high luminosity and eventual high fluence of energetic particles, 
detectors with fast response to fit the short shaping time of 20--25 ns and sufficient radiation hardness are required. 

Charge collection measurements carried out at the Ioffe Institute have shown a reversal of the pulse polarity in the detector response to short-range charge injection. Since the measured negative signal is about 30-60\% of the peak positive signal, the effect strongly reduces the CCE even in non-irradiated detectors. For further investigation of the phenomenon the measurements have been reproduced by TCAD simulations. 

As for the measurements, the simulation study was applied 
for the p-on-n strip detectors similar in geometry to those developed for the ATLAS experiment and for the Ioffe Institute designed p-on-n strip detectors with each strip having a window in the metallization covering the p$^+$ implant, allowing the generation of electron-hole pairs under the strip implant. Red laser scans across the strips and the interstrip gap with varying laser diameters and Si-SiO$_2$ interface charge densities ($Q_\textrm{f}$) were carried out. The results verify the experimentally observed negative response along the scan in the interstrip gap. When the laser spot is positioned on the strip p$^+$ implant the negative response vanishes and the collected charge at the active strip % proportionally 
increases respectively. 
 
The simulation results offer a further insight and understanding of the influence of the oxide charge density in the signal formation. The main result of the study is that a threshold value of $Q_\textrm{f}$, that enables negligible losses of collected charges, is defined.
The observed effects and details of the detector response for different charge injection positions are discussed in the context of Ramo's theorem.

\end{abstract}

\begin{keyword}
% \texttt{elsarticle.cls}\sep \LaTeX\sep Elsevier \sep template
% \MSC[2010] 00-01\sep  99-00
% Silicon detectors; Radiation hardness; Defect characterization; Detector modelling and simulations; Full Detector Systems
Silicon radiation detectors; Strip sensors; Transient current; Charge collection; TCAD simulations
\end{keyword}

\end{frontmatter}

% \linenumbers

% \section{The Elsevier article class}
\section{Introduction}
\label{Introduction}
Segmented silicon detectors, like micropixel and microstrip, are the primary detector types used in inner trackers of the Large Hadron Collider (LHC) experiments \cite{Robinson2002,Andricek2000,Allport2000}. Due to the high luminosity of the proton beam and the high counting rate the detectors with fast response to fit the short shaping time of 20--25 ns are required. The detectors must operate in a high-radiation environment for at least 10 years, maintaining a sufficient performance regardless of the accumulating degradation of the silicon properties.

The development and testing of advanced detectors include detailed studies of the detector current and charge responses using e.g. sub-nanosecond lasers, which % make possible 
enable to simulate the interaction of short or long range (MIP) particles with the detector. This gives information on the carrier transport in the detector bulk and the electric field distribution which are the input data for the prediction of the long-term scenario of the detector position sensitivity and charge collection efficiency (CCE) degradation. The latter is the governing parameter to define the guidelines of the detector operation after installation to the high energy physics experiments.

Earlier studies of silicon surface-barrier strip detectors have displayed a distinctive feature in the detector response - pulse responses of reversed polarity from the strips adjacent to the collecting strip, i.e. closest to the position of charge injection \cite{Kraner1983}. Later bipolar current pulse and negative charge signals from the adjacent strips were observed from non-irradiated planar Si strip detectors with a p-on-n structure \cite{Eremin2003,Verbit2005} where non-equilibrium carriers were generated via illumination of the detector strip side by a focused red laser pulse. The signals were recorded in the scanning mode and the negative charge signal reached up to 60\% of the positive one \cite{Verbit2005}. The results were qualitatively explained by formation of a potential minimum under the Si-SiO$_2$ interface due to the positive charge in the SiO$_2$ layer. Evidently, potential distribution inside this minimum, which acts as a sink for collected electrons, depends on the interface properties thus affecting the charge losses.

The effect of the signal reversal has a potential to strongly influence the interpretation of the results on charge collection in the detector bulk damaged by irradiation and can be also sensitive to the radiation damage of the SiO$_2$ layer and the interface. Recently the influence of these factors on charge and current signals was demonstrated in the study of radiation and environmental effects on silicon strip detectors \cite{Poehlsen2013,Poehlsen2013b,Poehlsen2013c}. It was shown that X-ray irradiation, biasing history and humidity led to reversal of the detector signal sign and incomplete charge collection and in a great extent affected the time of the signal relaxation to its steady-state value. Using numerical simulations, this effect was explained by a disturbance of steady state of the accumulation layer beneath the Si-SiO$_2$ interface in the interstrip gap.

The results presented below continue the study described in \cite{Eremin2003,Verbit2005} with the goal to quantify the influence of the charge accumulated at the Si-SiO$_2$ interface within the interstrip gap on the charge and current induction on the strips. This is performed via simulation using TCAD, which allows reconstructing the details of the potential and electric field distributions and the charge and current response of the strips. The results of the simulations are compared with the experimental data and their correlation is demonstrated. It is shown that there is a threshold interface charge density which initiates the appearance of reversed polarity signal at the neighboring strips and significant increase of charge losses at the readout strip.
\section{Observations on charge induction in strip detectors}
\label{Observation}
The main feature of silicon strip detectors performance studied here is the inversion of the charge signal on the strips neighboring to the one physically collecting the charge. The guideline for the charge signal inversion in strip detectors follows from the Ramo’s theorem \cite{Ramo1939}. It predicts the result of electrostatic induction produced by the moving charge in an arbitrary system of conductive electrodes placed in a non-conductive medium. According to the theorem, the charge $q_0$ generated by incident particles and drifting in the medium induces the current $\textrm{i}_i(t)$ on any electrode as
\begin{equation}\label{eq1}
\textrm{i}_i(t)=q_0\vec{v}_\textrm{dr}(t){\vec{E}_i}^*(t)
\end{equation}
or the charge
\begin{equation}\label{eq2}
\Delta\textrm{Q}=q_0\Delta\varphi^*,
\end{equation} 
where $\vec{v}_\textrm{dr}$ is the drift velocity and ${\vec{E}_i}^*$ the weighting electric field, $\Delta\textrm{Q}$ the change of the charge $\textrm{Q}$ induced on the readout electrode, and $\Delta\varphi^*$ the change of the weighting potential $\varphi^*$. The weighting field is defined in the Ramo’s theorem as the electric field created by the unit potential ($V^*=1$) applied to the readout electrode while all the other electrodes of the device remain grounded ($V^*=0$). 

Transfer of a constant point charge created at some electrode (1) to the collecting electrode (2) will induce on the latter the full charge $\textrm{Q}=q$ since  
\begin{equation}\label{eq3}
\int\Delta\varphi^*=V^*=1
\end{equation}  
along any trajectory of the charge $q$ started at electrode (1) and ended at electrode (2). The shape of $\textrm{i}_i(t)$ is defined by the charge trajectory and can be different. For example, the weighting field distribution for any p$^+$ strip depends on the width of the strip, its pitch and the detector thickness.  Thus, it reaches the maximum value at the strip due to a geometrical focusing effect. At the same time, the trajectory of drifting charge is defined by both the distribution of the space charge density and geometrical focusing effect. This can lead to the significant difference between the electric field lines governing the collected charge trajectory and the lines of weighting electric field. 

% however, the result $\textrm{Q}=q$ will be the same irrespective of the trajectory. 
% Simultaneously, a 
Obviously, the current is induced on all % other 
non-collecting electrodes simultaneously. However, the weighting potential difference between the electrode (1) and any other electrode (different from the collecting one) equals zero, $\int\Delta\varphi^*=0$, and so does the induced charge, $\textrm{Q}=0$. This means that induced current arising during carrier drift has bipolar shape in which positive and negative current parts compensate each other. % Obviously this result also satisfies to the condition of charge conservation in the system of electrodes.
Evidently, this result also satisfies to the condition of charge conservation in the system of electrodes, i.e. the total charge induced on all electrodes is $\textrm{Q}=q$ irrespective of the trajectory.   
 
The result can be simply applied to the charge collection in silicon strip detectors with a surface-barrier or planar structure \cite{Radeka1988,Lutz1999,He2001,Spieler2005}. They contain a non-conductive medium (space charge region (SCR)), and surface barrier or implanted p$^+$ strips and the n$^+$ back contact as conductive electrodes and the charge is collected via electron and hole drift inside the detector. The same logic as above predicts that only on the strip which physically collects the charge generated at any other strip or at the back contact the positive charge signal Q equals $q_0$ and a unipolar current response will be measured. All other strips show a bipolar current response and zero charge.  
   
This statement is valid when the drifting charge is independent on time, i.e. there is no charge loss. Negative charge signal can be observed in the case of carrier trapping on defects or inside the potential minimum near the Si-SiO$_2$ interface. In this case the charge on the collecting electrode will be smaller, i.e. $\textrm{Q}\textless{q_0}$ and the positive and negative parts of the induced current do not compensate each other. Thus, the collected charge on the neighboring strips differs from zero and is negative. 
     
Following these basic considerations, the experimental results in \cite{Eremin2003,Verbit2005} demonstrating the negative collected charge on the neighboring strips and the bipolar current responses were explained by carrier trapping only. The detectors used in \cite{Eremin2003,Verbit2005} were not irradiated and therefore carrier trapping in the bulk should be insignificant that speaks in favor of the proposed explanation based on the charge losses in the potential minimum near the detector surface in the interstrip gap \cite{Verbit2005}.  

\section{TCAD simulation procedure and set-up}
\label{TCAD Simulations}
Due to their versatility and ability to reproduce not only generic behavior but also absolute values of measured silicon sensor characteristics before and after radiation, the Technology Computer-Aided Design (TCAD) numerical simulations have been applied as the next investigation tool in further studies of the % aforementioned effect of 
reverse polarity charge signals. The reliability and accuracy of the TCAD simulations of the silicon strip detectors have already been documented in several publications and Ph.D. theses, including \cite{Poehlsen2013,Poehlsen2013b,Peltola2014,Bhardwaj2014,Peltola2015,Peltola2015r,Peltola2016p,Dalal2014} and \cite{Eber2013,Peltola2016}, respectively.

% TCAD simulations not only provide the numerical calculations of charge collection parameters like the electric field distribution, bulk silicon properties and radiation induced defects for strip detectors, but also allow to perform these calculations with respect to parameters like Si-SiO$_2$ interface charge density or laser diameter that would be next to impossible to realize experimentally. TCAD simulations are an invaluable tool in the study the electric fields inside a segmented detector, since these cannot be measured directly.
%
All simulations % presented 
in this paper were carried out using the Synopsys Sentaurus\footnote{http://www.synopsys.com} finite-element Technology Computer-Aided Design (TCAD) software framework. % In all modelled sensor structures the silicon bulk material was considered to have $\langle100\rangle$ crystal orientation. 
% Common AC boundary conditions used in AC simulation are Neumann boundary and oxide-semiconductor jump conditions carried over directly from DC simulation; Dirichlet boundary conditions for carrier densities where \textit{n} and \textit{p} at Ohmic contacts are $n=p=0$; and Dirichlet boundary conditions for AC potential at Ohmic contacts that are used to excite the system.
All presented simulations apply % Simulation applies %Common AC boundary conditions used in AC simulation are 
Neumann boundary and oxide-semiconductor jump conditions (potential jump across the material interface due to dipole layers of immobile charges), % carried over directly from DC simulation; 
as well as Dirichlet boundary conditions for carrier densities % where \textit{n} and \textit{p} at Ohmic contacts are $n=p=0$;
and % Dirichlet boundary conditions for 
AC potential at Ohmic contacts that are used to excite the system. 

\begin{figure}[tbp]
     \centering
     \subfloat[ATLAS sensor.]{\includegraphics[width=.7\textwidth]{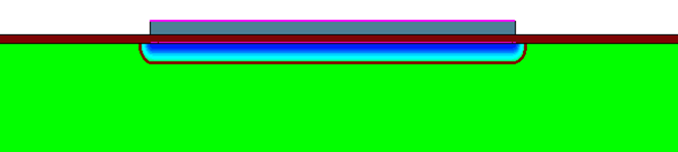}\label{ATLAS1}}\\
     \subfloat[Strip sensor modified from ATLAS-design.]{\includegraphics[width=.7\textwidth]{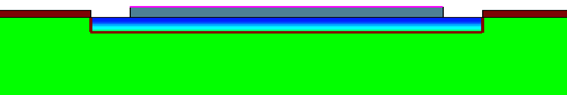}\label{ATLAS}}\\
     \subfloat[Ioffe Institute sensor.]{\includegraphics[width=.7\textwidth]{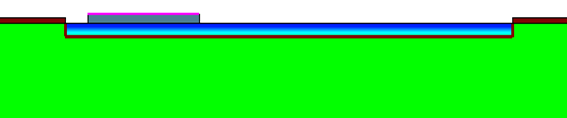}\label{PTI}}
    \caption{One strip close-up of the simulated 2-dim. 3-strip p-on-n sensor structures with parameters from table~\ref{tab:2}. % Both (\ref{ATLAS}) modified ATLAS and (\ref{PTI}) Ioffe Institute "baby" strip detectors were 300 $\mu\textrm{m}$ thick with a metal overhang of -2 $\mu\textrm{m}$. 
The (\ref{ATLAS1}) ATLAS strip sensor is capacitively coupled while both (\ref{ATLAS}) modified ATLAS sensor design and (\ref{PTI}) Ioffe Institute % "baby" strip detectors were 300 $\mu\textrm{m}$ thick with a metal overhang of -2 $\mu\textrm{m}$. 
sensors are DC-coupled. The window in the metallization of the Ioffe Institute detector % has also a 
for a charge injection from oxide-free region is visible in (\ref{PTI}). 
The Al layers on top of the structures are illustrated in gray and the surrounding oxide layers are shown in brown. Strip implants are shown in blue 
while the green region is the lightly doped Si bulk. The n$^+$ layer and the metallization of the non-segmented backplane are not pictured.}
\label{fig:8b}
\end{figure}

% Since the charges within the oxide layer interact with the active area of the silicon via the Si-SiO$_2$ interface, as an effective approach their presence was modelled solely by the interface charge density ($Q_\textrm{f}$). Recombination throughout the silicon was implemented by recombination through deep defect levels in the bandgap, % is usually labeled i.e. Shockley-Read-Hall (SRH) recombination, and band-to-band Auger recombination. Initial simulations with additional surface SRH recombination resulted in decreased reverse polarity signals at charge injection points beyond midgap from the collecting strip (e.g. at $Q_\textrm{f}=1.3\cdot10^{11}~\textrm{cm}^{-2}$ the maximum collected reverse polarity charge for surface SRH with intial recombination velocities set to $10^{4}~\textrm{cm/s}$ was 57\% from the corresponding charge without surface SRH), moving the simulated results further away from experimentally observed behavior \cite{Eremin2003,Verbit2005}. Thus, the simulations presented in section~\ref{Results} do not include surface SRH recombination. Since interface trap density in non-irradiated silicon strip sensor has been reported to be about two orders of magnitude lower than interface charge density \cite{Poehlsen2013}, for simplicity these were not considered in the simulations.
%%%%%%%%%%%%%%%%%%%%%%%%%%%%%%%%%%%%%%%%%%%%%%%%%%%%%%%%%%%%%%%%%%
% \subsection{Simulation set-up}
% \label{Simulation set-up}
For the simulation study of the observed charge collection behaviour in % figure~\ref{fig:fig4} 
Refs.~\cite{Eremin2003,Verbit2005} the 2-dimensional structures presented in figure~\ref{fig:8b} were applied. A two dimensional simulation
was considered to be
% This was % deemed necessary 
% considered 
sufficient approach since, unlike for pixels, strip segmentation is monotonous along the second coordinate parallel to the sensor surface and provides no additional contribution to the local electric field evolution with voltage at the strip edges.
% in a 2-dim. structure only interactions between a single column of pixels would be monitored. Also the correct reproduction of the local electric fields due to the circular shape of the pixels required a 3-dim. structure. Since most of the pixels in the detector do not experience any influence from the active edges, these were not included in the simulated structure, as can be seen from figure~\ref{fig:8b}.
% 
% For the simulations of the p-on-n and p-on-p active edge pixel sensors a 5 pixel or larger structure was used to avoid any non-uniformities from the border effects on the mesh formation at the center part of the device where the position of the charge injection was set. % Presented in figure~\ref{fig:8} is the principle of the 2-dimensional device structure for the two sensor types. 
% As can be seen, the heavily doped phosphorus (P) implantations (n$^+$) surround the device bulk from three sides while the pixels were constructed by the heavily doped boron (B) implantations (p$^+$) and aluminum contacts on the front surface. The p-on-n and p-on-p devices were simply generated by applying light P or B doping to the substrate, respectively. 

% All modelled strip sensor configurations, ATLAS and Ioffe Institute, had 
The physical thicknesses, pitches, strip widths and layer dimensions of the modelled strip sensor configurations are given in table~\ref{tab:2}. The comparison between experimental results and simulations were carried out on sensor designs in figures~\ref{ATLAS1} and~\ref{PTI} with parameters as close to the real sensors as possible (AC-coupled strips, thickness 285 $\mu\textrm{m}$ and resistivity 3--4 k$\Omega\cdot\textrm{cm}$ for the ATLAS and DC-coupled strips, thickness 300 $\mu\textrm{m}$ and resistivity 5 k$\Omega\cdot\textrm{cm}$ for the Ioffe Institute design, respectively \cite{Eremin2003,Verbit2005}).
% All modelled strip sensor configurations, ATLAS and Ioffe Institute, % were designed with parameters as close to the real sensors as possible. Both sensor types 
% had physical thicknesses set to 300 $\mu\textrm{m}$, while the pitches, strip widths and  
% a pitch of 55 $\mu\textrm{m}$ and a pixel implant diameter %, i.e. width in 2-dim. structure, of 30 $\mu\textrm{m}$. 
% layer dimensions % and doping parameters 
% are given in table~\ref{tab:2}. % in which 
For a simulation investigation focused on the differences in strip configurations, a modified ATLAS-design strip sensor structure (figure~\ref{ATLAS} and table~\ref{tab:2}), with equal physical thickness, strip coupling and interstrip gap to Ioffe Institute design, was applied. 
% This was done to minimize any effect from the small discrepancies in thickness and resistivity in the real sensors (285 $\mu\textrm{m}$, 3--4 k$\Omega\cdot\textrm{cm}$ for the ATLAS and 300 $\mu\textrm{m}$, 5 k$\Omega\cdot\textrm{cm}$ for the Ioffe Institute design, respectively \cite{Eremin2003,Verbit2005}) and focus the simulation investigation to the differences in strip configurations. 
Also the bulk dopings % were estimated by using the resistivity data 
of the two sensor substrates were set equal $N_\textrm{B}=1\times10^{12}~\textrm{cm}^{-3}$, that resulted in full depletion voltages well below 100 V. % The lateral diffusion of the strip implants was set to 0.8$\times$depth. % The aluminum metallizations above the pixel implants and their vias through the oxide layer had the diameters of 36 $\mu\textrm{m}$ and 24 $\mu\textrm{m}$, respectively. Detailed cross-sectional slice of the pixel is presented in figure~\ref{fig:8c}. Since the only differences between the sensor types (both had p$^+$ implantations at the pixels and n$^+$ at the non-segmented side) were the pixel implant diffusion depths and the type and concentration of the bulk doping, the figures~\ref{fig:8b} and~\ref{fig:8c} can be considered to represent both of the simulated sensor structures. % It should be noted that in the figure~\ref{pixel5} the pitch has a diagonal orientation while in figure~\ref{pixel9} it is parallel to the sides.% The gap between the device edge and first/last pixel was set to 50 $\mu\textrm{m}$. % For the n-on-p sensors the p-stop widths were 4 $\mu\textrm{m}$ and the spacing between p-stops was 6 $\mu\textrm{m}$, while the peak doping of $1\times10^{16}$ $\textrm{cm}^{-3}$ decayed to the bulk doping level within 1.5 $\mu\textrm{m}$, having equal depth with n$^+$ strip implants. 
% The thicknesses of the oxide on the front surface and Al layers on the front and backplanes were set to 250 nm and 500 nm, respectively. 
Lateral diffusion of the strip implant, that results in somewhat wider strip implant than its metallization when the two have equal nominal widths \cite{Eremin2003}, was approximated by % using smaller width
using Al width that left its edge 2 $\mu\textrm{m}$ from the implant edge, as displayed in figures~\ref{ATLAS} and~\ref{PTI}.  

The peak concentrations of both backplane n$^+$ and strip p$^+$ implantations were $5\times10^{18}$ $\textrm{cm}^{-3}$ while both decayed to the bulk doping level within 1.0 $\mu\textrm{m}$ depth using a gaussian profile. 
% The area of the real "baby" strip detectors
The area of the strip detectors used in measurements, 10$\times$10 mm$^2$, was taken into account by stretching the 2-dimensional structures to the third dimension by an area factor. % As the earlier measurements showed the density of the interface states to be about $Q_\textrm{f}=5\cdot10^{10}~\textrm{cm}^{-2}$ \cite{Verbit2005}, this was used as the baseline for the study of the effect of varied $Q_\textrm{f}$ at the Si-SiO$_2$ interface. 

As the % reported 
density of the interface states observed in the Ioffe Institute-design strip sensors was typically % about 
in the range of $Q_\textrm{f}=5-8\cdot10^{10}~\textrm{cm}^{-2}$ % \cite{Verbit2005} (also up to $8\cdot10^{10}~\textrm{cm}^{-2}$)
% As the % earlier measurements showed the 
% reported density of the interface states in the Ioffe Institute-design strip sensor was about $Q_\textrm{f}=5\cdot10^{10}~\textrm{cm}^{-2}$ \cite{Verbit2005} (also up to $8\cdot10^{10}~\textrm{cm}^{-2}$) 
and in ATLAS-design strip sensors about $2\cdot10^{11}~\textrm{cm}^{-2}$ \cite{Robinson2002}, these were included % was used as the baseline for the study of the effect of varied 
within the range of studied $Q_\textrm{f}$ values at the Si-SiO$_2$ interface.

All modelled sensors consisted of 3-strip structures, of which one strip close-ups are presented in figure~\ref{fig:8b}. Illustrated in figure~\ref{PTI} is the Ioffe Institute design's % each strip has a 
window in metallization % that covers 
of the p$^+$ implant % . The important advantage of this design is that it 
that allows the generation of electron-hole (\textit{e--h}) pairs under the strip implant. % As can be seen from figure~\ref{fig:8b}, the metal overhangs of the strip implants in both sensor designs were set to -2 $\mu\textrm{m}$ to reduce the differences in strip metallizations to the Ioffe Institute design's charge injection window. % Each strip had a DC-coupled electrode at zero potential. The readout electrode was the centermost strip of the 3-strip sensor in all simulations. The reverse bias voltage was provided from the backplane n$^+$ contact.
% To gain a more complete picture of the negative polarity signal dependencies, the laser scans were extended to a n-on-p strip sensor with otherwise identical strip and bulk parameters to the PTI sensor design. Two configurations of the n-on-p sensor were applied, strips isolated by a p-spray implantation, presented in figure~\ref{fig:8c}, and without any strip isolation.
%
The readout electrode was the centermost strip % contact of a detector 
of the 3-strip sensor structure in all simulations. In the figures of section~\ref{Results} it is always located at $x=0~\mu\textrm{m}$. 
The reverse bias voltage was provided from the backplane contact. In all designs each strip had a DC-coupled electrode at zero potential. For designs in figures~\ref{ATLAS} and~\ref{PTI} this was also the electrode used for charge collection, while for ATLAS-design in figure~\ref{ATLAS1} charge collection was done by a capacitively coupled electrode separated from the silicon bulk by the oxide layer.

To generate charge carriers in the detector, a red laser with a % pulse length of 1 ns, 
wavelength of 670 nm and penetration of about 7 $\mu\textrm{m}$ in silicon, was applied. The pulse length was set to 1 ns to match the experimental value in Refs. \cite{Eremin2003,Verbit2005}. With the intensity set to 20 W/cm$^2$, this resulted in laser generated \textit{e--h} pair density of about 345 $\mu\textrm{m}^{-2}$. % Presented in figure~\ref{fig:8d}, 
Two laser spot diameters were used in the simulations, a 10 $\mu\textrm{m}$ which was also the diameter applied in the measurement set-up \cite{Eremin2003,Verbit2005,Ruggiero2003} and a 1 $\mu\textrm{m}$ % to achieve a scanning resolution beyond experimentally possible resolution\\ \\
to enable scanning beyond experimentally available resolution. % The 1 $\mu\textrm{m}$ laser spot diameter was also applied to % scans with 
% an infrared laser (IR) with a wavelength of 1060 nm and a penetration depth of several hundred microns. % When all simulated illuminations we done from the strip-side of the sensor, 
The red % and IR 
laser allows % both
% one to 
the investigation of a % separate collection of holes and 
transient signal generated separately by the drift of electrons or holes (depending on which side of the detector is being illuminated) % and the response of the detector to a charge injection similar to the one produced by minimum ionizing particles, respectively 
\cite{Sze1981}. All laser scans were carried out at room temperature ($T=293~\textrm{K}$). 
\begin{table}[tbp]
\caption{Active thicknesses, layer dimensions and strip parameters for the three simulated sensor structures. The $W_{\textrm{impl,Al}}$ are the widths of the strip implant and its metallization, respectively. % Three-level defect model included in the non-uniform 3-level model%\cite{bib2}
%, parametrized for the fluence range $(0.3 - 1.5)\times10^{15}$ n$_{\textrm{\tiny eq}}\textrm{cm}^{-2}$. 
The $t_{\textrm{Ox,Al}}$ are the oxide and aluminum layer thicknesses, respectively, while $d_{\textrm{impl}}$ is the depth of the strip implantation.} % $\sigma$$_{\textnormal{\tiny e,h}}$ are the electron and hole capture cross sections and $\Phi$ is the fluence.
% The depths and peak concentrations for the Gaussian decay of the heavily doped implantations to the bulk level are also given.}{\small {\bf $W_{\textnormal{\tiny Al}}$}}
\label{tab:2}
\smallskip
\centering
\begin{tabular}{|c|ccccccc|}
    \hline
    {\bf Sensor type} & \multicolumn{1}{c|}{Thickness} & \multicolumn{1}{c|}{Pitch} & \multicolumn{1}{c|}{\bf $W_\textnormal{impl}$} & \multicolumn{1}{c|}{\bf $W_\textnormal{Al}$} & \multicolumn{1}{c|}{\bf $t_{\textnormal{Ox}}$} & \multicolumn{1}{c|}{\bf $t_{\textnormal{Al}}$} & {\bf $d_{\textnormal{impl}}$}\\
     & \multicolumn{1}{c|}{\small \textnormal{[$\mu\textrm{m}$]}} & \multicolumn{1}{c|}{\small \textnormal{[$\mu\textrm{m}$]}} & \multicolumn{1}{c|}{\small \textnormal{[$\mu\textrm{m}$]}} & \multicolumn{1}{c|}{\small \textnormal{[$\mu\textrm{m}$]}} & \multicolumn{1}{c|}{\small \textnormal{[$\mu\textrm{m}$]}} & \multicolumn{1}{c|}{\small \textnormal{[$\mu\textrm{m}$]}} & {\small \textnormal{[$\mu\textrm{m}$]}}\\
    \hline
    ATLAS & 285.0 & 80.0 & 18.0 & 18.0 & 0.47 & 0.70 & 1.0\\
    \hline
    Modified ATLAS & 300.0 & 80.0 & 20.0 & 16.0 & 0.47 & 0.70 & 1.0\\
    \hline
%    {\small PTI} & {\small 100.0} & {\small 40.0} & {\small 10.0} & {\small 0.47} & {\small 0.70} & {\small 1.0}\\
    Ioffe Institute & 300.0 & 100.0 & 40.0 & 10.0 & 0.47 & 0.70 & 1.0\\
%    {\small Shallow acceptor} & {\small $E_{\textnormal{c}}-0.40$} & {\small 8$\times$10$^{-15}$} & {\small 2$\times$10$^{-14}$} & {\small 14.417$\times$$\Phi$+3.168$\times$10$^{16}$}\\
    \hline
\end{tabular}
\end{table}

\section{Simulation results}
\label{Results}
The simulated charge collection plots resulting from the red laser interstrip scans presented in this section, are always normalized to the maximum charge collected at the lowest value of Si-SiO$_2$ interface charge density $Q_\textrm{f}$ in each figure. This is to provide straightforward comparison of varied values of $Q_\textrm{f}$ and laser diameter for the given strip sensor configuration.

Since the charges within the oxide layer interact with the active area of the silicon via the Si-SiO$_2$ interface, as an effective approach their presence was modelled solely by the interface charge density ($Q_\textrm{f}$). Recombination throughout the silicon was implemented by recombination through deep defect levels in the bandgap, % is usually labeled 
i.e. Shockley-Read-Hall (SRH) recombination \cite{Shockley1952}, and band-to-band Auger recombination \cite{Burhop1952}. The doping dependence of SRH recombination resulted in non-uniform recombination lifetime distribution, % in silicon, 
reducing the recombination lifetimes from bulk values % with about 
up to three orders of magnitude % shorter lifetimes at the strip implant than in the bulk. 
in the vicinity of heavily doped regions of strip implants and backplane blocking contact. The bulk recombination lifetimes, 100 ms for electrons and 10 ms for holes, were tuned according to an earlier % simulation 
study \cite{Bhardwaj2014}, where simulated carrier lifetimes were tuned to reproduce the measured low leakage currents of non-irradiated p-on-n detectors.

% Initial simulations with additional surface SRH recombination resulted in decreased reverse polarity signals at charge injection points beyond midgap from the collecting strip (e.g. at $Q_\textrm{f}=1.3\cdot10^{11}~\textrm{cm}^{-2}$ the maximum collected reverse polarity charge for surface SRH with intial recombination velocities set to $10^{4}~\textrm{cm/s}$ was 57\% from the corresponding charge without surface SRH), moving the simulated results further away from experimentally observed behavior seen in figures~\ref{atlasM} and~\ref{fig:fig13}. Thus, the simulations presented in this section do not include surface SRH recombination. Since interface trap density in non-irradiated silicon strip sensor has been reported to be about two orders of magnitude lower than interface charge density \cite{Poehlsen2013}, for simplicity these were not considered in the simulations.
%
%%%%%%%%%%%%%%%%%%%%%%%%%%%%%%%
As stated above, TCAD package allows considering recombination in the silicon detector as a combination of % the above mentioned %recombination through the deep defect levels in the bandgap (i.e. Shockley-Read-Hall (
SRH mechanism and % band-to-band 
Auger recombination. However, in the space charge region of the detector the SRH recombination is inefficient since a steady-state concentration of free carriers is negligible. Auger recombination can be neglected as well since the carrier concentration generated by laser pulses is too low. Therefore only carrier trapping can be a factor reducing the collected charge. However, this study was linked to the experiments performed on non-irradiated Si detectors where trapping in the detector bulk is negligible (typical trapping time is in the range of milliseconds leading to charge losses at the level of ratio $t_\textrm{coll}/\tau_\textrm{tr}=10^{-8}~\textrm{s}/10^{-3}~\textrm{s}=10^{-5}$, where $t_\textrm{coll}$ is the charge collection time and $\tau_\textrm{tr}$ is the carrier trapping time). %, i.e. recombination lifetime.).

The surface recombination was estimated based on the results of \cite{Poehlsen2013} where it was shown that the density of surface traps in the interstrip gap of non-irradiated silicon sensor is about two orders of magnitude lower than the interface charge density. Thus, for $Q_\textrm{f}=1\times10^{11}~\textrm{cm}^{-2}$ the density of surface traps should be in the order of $10^{9}~\textrm{cm}^{-2}$. Estimation of the surface recombination velocity using a rather large capture cross-section of surface trapping centers of $1\times10^{-14}~\textrm{cm}^{-2}$ and thermal velocity at room temperature of $2\times10^{7}~\textrm{cm/s}$ \cite{Sze1981} gives the value of 200 cm/s. Since this value does not affect the accuracy of the treatment of the experimental data used in this study, for simplicity the simulations presented below do not consider the surface SRH recombination.
%%%%%%%%%%%%%%%%%%%%%%%%%%%%%%%%

Since the oxide and interface charges depend on the silicon wafer orientation, being larger for $\langle111\rangle$ wafers \cite{Zhang2012}, in the simulation the interface charge was varied within $0.7\cdot10^{10}-4\cdot10^{11}~\textrm{cm}^{-2}$ covering the range for both $\langle100\rangle$ and $\langle111\rangle$ orientations \cite{Robinson2002,Verbit2005,Poehlsen2013,Moscatelli2016}.
% \subsection{Simulation results}
% \label{Simulation results}
%%%%%%%%%%%%%%%%%%%%%%%%%%%
%%%%%%%%%%%%%%%%%%%%%%%%%%%
\subsection{Interstrip scans with 10 $\mu\textrm{m}$ diameter red laser}
\label{10um}
\begin{figure}[tbp] 
\centering
 \subfloat[Measured and simulated red laser scans of ATLAS strip sensor.]{\includegraphics[width=.65\textwidth]{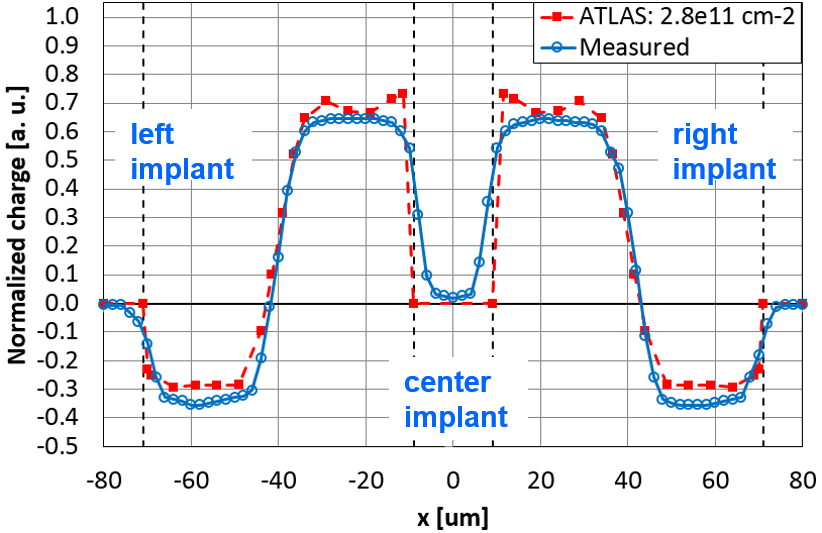}\label{atlasM}}\\
 \subfloat[Red laser scans with varied $Q_\textrm{f}$.]{\includegraphics[width=.65\textwidth]{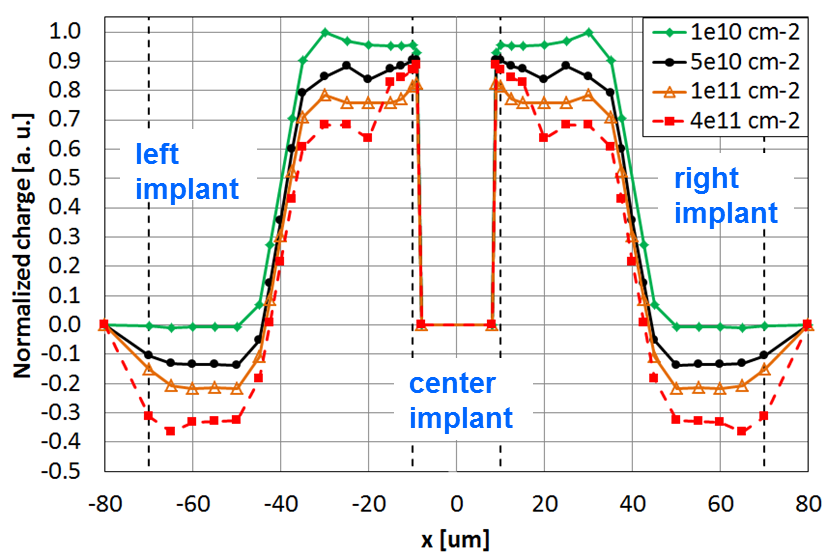}\label{atlas10}}
\caption{% Charges collected at the centermost strip of the ATLAS strip detector presented in figure~\ref{ATLAS1} after a red laser scan across three strips for (a) measured and simulated sensor 
Charges collected at the centermost strip after a red laser scan across the three strips of (a) measured and simulated ATLAS strip detector presented in figure~\ref{ATLAS1} and for (b) varied values of interface charge density $Q_\textrm{f}$ in a simulated strip sensor structure modified from ATLAS-design presented in figure~\ref{ATLAS}. % that is in units of cm$^{-2}$. 
The bias voltages during the charge collection were (a) 350 V and (b) 300 V while the laser diameter was 10 $\mu\textrm{m}$. The strip implant edges are marked by dashed black lines. The simulated charges in (b) are normalized to the maximum charge collected at $Q_\textrm{f}=1\times10^{10}~\textrm{cm}^{-2}$. The measured curve in (a) is from Ref. \cite{Eremin2003}.}
\label{fig:fig11}
\end{figure}

Simulation results for 
the cross strip scans of a strip detector with ATLAS design from the p$^+$-strip side show that the collected charge correlates well with the topology of the
strips and the 80 $\mu\textrm{m}$ pitch is clearly recognizable, as can be seen from figure~\ref{fig:fig11}. 
%
% As can be seen from figure~\ref{fig:fig11}, 
% In 
% the cross strip scans of a strip detector with ATLAS design % measured 
% with the red laser illuminating 
% from the p$^+$-strip side show that % presented in figure~\ref{fig:fig11},  
% the collected charge correlates well with the topology of the strips and the 80 $\mu\textrm{m}$ pitch is clearly recognizable. % distinguished. 
% Moving from the 'left implant' towards 
When the laser position is either at the 'center implant', which is considered as the readout strip, or its adjacent strips,  
the resulting plots for the collected charge ($Q_\textrm{coll}$) % for the p$^+$-strip side 
contain low signal regions due to the opacity of the % 16 $\mu\textrm{m}$ 
strip implant wide metallization.
The simulated laser produces optical carrier generation that has parallel to the sensor surface a Gaussian distribution within 10 $\mu\textrm{m}$. 
Thus, if the peak value of optical carrier generation is minimized by the Al layer, also the tails of the distribution are minimized. % This leads to step function-like behavior of the simulated charge collection curves seen in figure~\ref{fig:fig11} when the laser spot is moved above the edge of the Al layer.
This leads to step function-like behavior of the simulated charge collection curves at the edge of the strip metallization that is not seen in the measured curve in figure~\ref{atlasM}. % when the laser spot is moved above the edge of the Al layer.

% The signal distribution % with the laser on the p$^+$-side shows that the charge sharing % is lower than moves from 50\% to about 20\% with increased values of $Q_\textrm{f}$ at the midpoint between strips. This means that when the charge is injected at the midgap, only about 40\% of the maximum collected charge is recovered by the two surrounding strips at the highest value of $Q_\textrm{f}$. 
As in earlier measurements \cite{Eremin2003,Verbit2005}, a significant % negative collected charge 
reversed polarity signal is observed % . This negative signal appears in the case of
when the point of carrier injection is % on the opposite side of the interstrip gap.
beyond midgap distance from the readout strip, i.e. when % $x\textless80~\mu\textrm{m}$ and 
$\lvert x\rvert\textgreater40~\mu\textrm{m}$ in figure~\ref{fig:fig11}. 
% If the value at the implant edge of the adjacent strip to the readout strip is excluded, 
% The % negative 
% reverse polarity signal around $\lvert x\rvert=50~\mu\textrm{m}$ varies as a function of $Q_\textrm{f}$ between 1\% and 33\% of the positive maximum, while the corresponding measured value is above 50\%, as shown in figure~\ref{atlasM}. The measured location of polarity reversal and the slope of the charge collection curve around midgap are closely reproduced by the simulation.
The reverse polarity signal around $\lvert x\rvert=50~\mu\textrm{m}$ and the positive signals between $\lvert x\rvert=20-30~\mu\textrm{m}$ % varies as a function of 
for $Q_\textrm{f}=2.8\times10^{11}~\textrm{cm}^{-2}$ % between 1\% and 33\% of the positive maximum, while 
are within 6\% and 2--7\%, respectively, from the corresponding measured values, % while the positive signal at $\lvert x\rvert=20~\mu\textrm{m}$ % is above 50\%, 
as shown in figure~\ref{atlasM}. The measured location of polarity reversal and the slope of the charge collection curve around midgap are closely reproduced by the simulation.

The normalization of the measured charge collection curve in figure~\ref{atlasM} was carried out so, that the sum of the absolute maximum values of the positive and negative polarity charges % collected at distances less and beyond of the interstrip midgap from the collecting strip, was 
would be equal to one.
% The normalization of the measured charge collection curve in figure~\ref{atlasM} was carried out so that the sum of the absolute maximum values of the charges collected at distances less and beyond of the interstrip midgap from the collecting strip, was equal to one. 
As can be seen from figure~\ref{atlas10}, this is also the consequence of normalization to the maximal simulated collected charge for the lowest value of $Q_\textrm{f}$ that has negligible contribution from the reverse polarity signals (and thus, undiminished positive polarity signals at distances less than midgap from the collecting strip). % for the simulated curves. 
% As can be seen from figure~\ref{atlas10}, this is also the consequence of normalization to the maximal collected charge for the lowest value of $Q_\textrm{f}$ that has negligible contribution from the reverse polarity signals for the simulated curves. 
These approaches enable reliable comparison between the measured and simulated charge collection curves. 

The charge collection scans display a strong dependence on interface charge density $Q_\textrm{f}$. When $x\textgreater$~midgap from the centermost strip in figure~\ref{atlas10}, the highest value of $Q_\textrm{f}$ corresponds to the highest negative collected charge. However, the effect essentially vanishes at the lowest $Q_\textrm{f}$ value of $1\times10^{10}~\textrm{cm}^{-2}$. When $x\textless$~midgap, reversed behavior is observed, the highest $Q_\textrm{f}$ now corresponds to the lowest positive collected charge until the implant edge is reached. The two highest value $Q_\textrm{f}$ plots also present sensitivity to the strip implant edges. The electric field peaks at the implant edges react strongly to the increase of $Q_\textrm{f}$ and start to amplify charges, both positive and negative, injected at their position. This is especially apparent for the highest $Q_\textrm{f}$ value $4\times10^{11}~\textrm{cm}^{-2}$, which around 5 $\mu\textrm{m}$ from the implant edge of the collecting strip surpasses in charge collection the $1\times10^{11}~\textrm{cm}^{-2}$ curve and converges with the charge collection results of $Q_\textrm{f}=5\times10^{10}~\textrm{cm}^{-2}$. 
More detailed investigation of the effects of charge multiplication after a charge injection at the implant edge follows in section~\ref{Backplane}.% more than 50\% of the positive one and is not much affected by the applied bias.

When the simulations were repeated for an otherwise identical strip sensor structure but with either DC- (direct contact to strip implant) or capacitively-coupled (oxide layer between the strip implant and collecting contact) charge collection electrodes, 
% Since the silicon sensors at the LHC-experiments are typically AC-coupled, a strip sensor configuration with AC-coupled strips was also investigated. Capacitive coupling was realized by providing the isolating oxide layer between the strip implant and collecting electrode in the Ioffe Institute sensor structure shown in figure~\ref{PTI}. The agreement of charge collection results between DC- and AC-coupled sensors after simulated red laser scans was within 2.5\% at any laser position, thus displaying an essentially coupling independent behavior.
the results of red laser scans were within 2.5\% at any laser position, thus displaying an essentially coupling independent behavior.

\begin{figure}[tbp] 
\centering
\includegraphics[width=.65\textwidth]{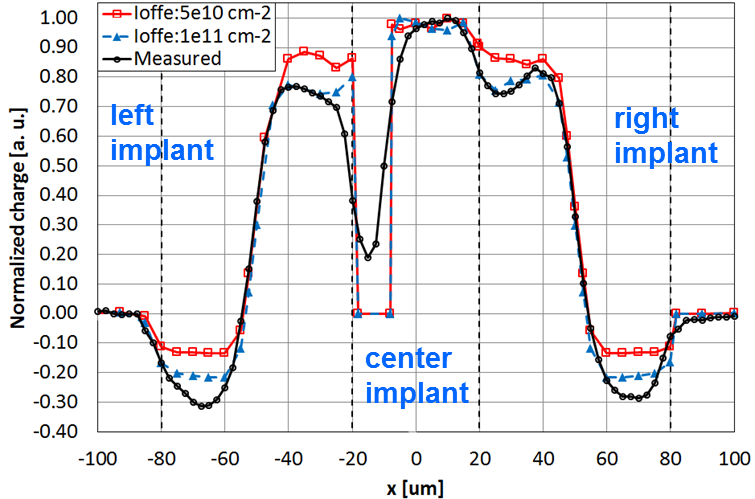}
\caption{Charges collected at the centermost strip of the Ioffe Institute-design strip detector presented in figure~\ref{PTI} after a red laser scan across three strips for two values of interface charge density $Q_\textrm{f}$. The charge collection was done at $V=400$ V while the laser diameter was 10 $\mu\textrm{m}$. The strip implant edges are marked by dashed black lines. % The simulated charges are normalized to the maximum charge collected at $Q_\textrm{f}=5\times10^{10}~\textrm{cm}^{-2}$. 
Both the measured and simulated charges are normalized to the maximum charge collected at % $Q_\textrm{f}=5\times10^{10}~\textrm{cm}^{-2}$
the center strip implant, free of oxide layer. The measurement is from Ref. \cite{Verbit2005}.}
\label{fig:fig13}
\end{figure}

% The signal distribution measured with the infrared laser correlates well with the one observed with the red laser. A scan of the p$^+$-side between the Al strips does not show any peculiarities, corresponding to a uniform pair generation through the whole bulk. The inverted polarity signal is not observed when the light spot is positioned at a distance beyond the midgap from the collecting strip.
%
% The features of the collected charge distribution vs. position described above are also seen in the current responses %. Note here that the current response is
% \begin{equation}%\label{eq.1}
% i(t)=\frac{\textrm{d}}{\textrm{d}t}Q_{\textrm{coll}}(t).
% \end{equation}
% For red laser injection on the p$^+$-side in the area close to the Al metallization the positive charge collected at the centermost strip corresponds to a positive polarity current pulse with a gradual decay. In the interstrip gap the current response decreases in amplitude and then inverts polarity at the midgap. The amplitude of the negative pulse then increases, reaching a maximum at the border of the p$^+$-implant of the adjacent strip. The duration of the negative current response is close to the width of the positive current response.
The corresponding measured and simulated red laser scans % of the laser, which illuminates the p$^+$ side of the 
of a detector with Ioffe Institute design % for operational bias of 400 V that is far above the full depletion voltage, 
are presented in figure~\ref{fig:fig13}. The scans are now carried out across the strips with implant width of 40 $\mu\textrm{m}$ but with metallization of only 10 $\mu\textrm{m}$, as presented in figure~\ref{PTI}. 
Scan results in the SiO$_2$ layer region and at the strip impants adjacent to the readout strip at the center show similar behavior to the % results of 
ATLAS-design sensor results in figure~\ref{fig:fig11}, while the drop of the signal from centermost strip now occurs only within the area of the metallization window % at the left part of the % when the laser spot is within the metal layer over the 
of the p$^+$ implant. % where the narrow metallization is located.

As seen in figure~\ref{fig:fig13}, both in measurement and simulation the maximum collected charges are observed in the area free from the metal contact at the centermost implant ($x$ coordinates between -20--20 $\mu\textrm{m}$). The normalization of both experimental and simulated curves is now done by setting the highest charge collected at the implant's metallization window, where no charge losses occur, equal to one. Complete charge collection at the metallization window is also reflected in $Q_\textrm{f}$ independent collection of the simulated charges, which is shown in greater detail in figure~\ref{S9_2}.   
% As in earlier measurements \cite{Verbit2005}, the maximum collected charges are observed in the area free from the metal contact at the centermost implant ($x$ coordinates between -20--20 $\mu\textrm{m}$). 
The amplitudes of the positive signals collected in the area of the oxide, with laser distance from the readout strip less than midgap, move from about 87\% to 74\% with increased values of $Q_\textrm{f}$, % the % signal is constant and the CCE is close to 100\%. 
% As in ATLAS detector scan, a negative charge response is observed along the laser scan at distances beyond midgap from the collecting strip. As soon as the light spot is placed on the p$^+$ implant of either one of the adjacent strips the negative response collected at the centermost strip disappears. 
%
% Hence the position of the negative signal well correlates to the region beyond the midgap from the collecting strip. 
% The charge sharing at the midgap moves from about 35\% to 30\% with increased $Q_\textrm{f}$. 
% The amplitude of 
while the negative signals at the region beyond the midgap from the collecting strip for different values of $Q_\textrm{f}$ % now varies 
vary between 13\%--22\% % (as opposed to the measured 30\%--40\% \cite{Verbit2005}) 
of % that of a 
the maximal positive signal. The simulated charge collection curves for the higher value of $Q_\textrm{f}$ are in close agreement with measured results at distances less or equal to midgap from the collecting strip, while the reversed polarity signals around $\lvert x\rvert=70~\mu\textrm{m}$ are within 7--10\% from measured values. % that is read out from the p$^+$ implant in the area free from a metal contact. 
% A sharp narrow drop in the signal corresponds to the scanning of the laser spot over a thin Al contact within the area of the metallization window with a width of 10 $\mu\textrm{m}$. % Since the diameter of the laser spot is larger than the scan step the experimental curves are "smoothed". 
% The lateral resolution of about 10 $\mu\textrm{m}$ is related to the diameter of the laser spot.

\begin{figure}[tbp] 
\centering
\includegraphics[width=.65\textwidth]{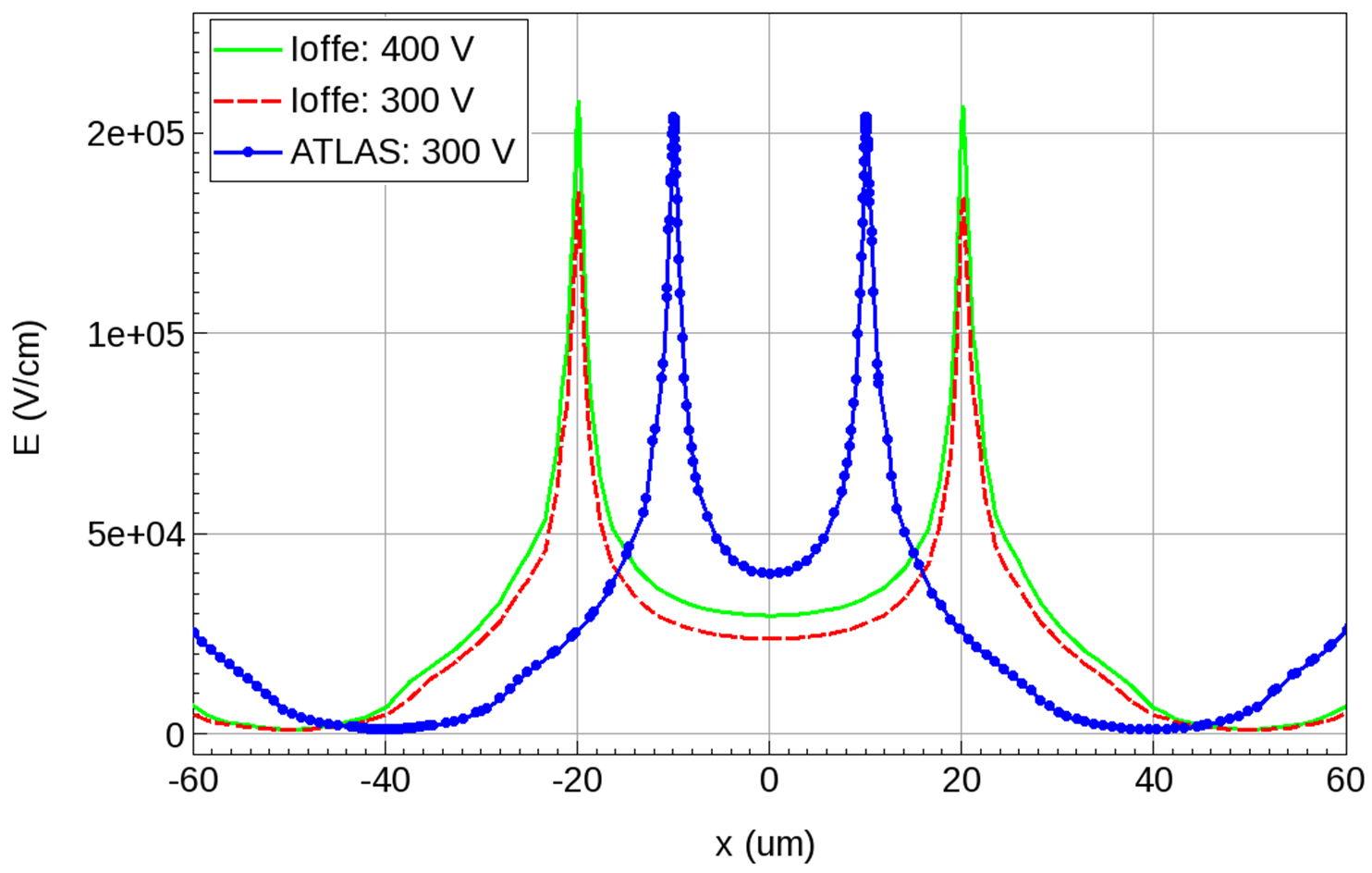}
\caption{Electric field distributions parallel to the detector surface at 0.8 $\mu\textrm{m}$ depth for the two detector designs in figures~\ref{ATLAS} and~\ref{PTI} with $Q_\textrm{f}=1\times10^{11}~\textrm{cm}^{-2}$. Voltages used for charge collection in both detectors are included, as well as $V=300$ V for the Ioffe Institute detector for clearer comparison with the strip detector modified from ATLAS-design.}
\label{fig:fig13b}
\end{figure}

When the charge collection plots of % $Q_\textrm{f}=1\times10^{11}~\textrm{cm}^{-2}$ 
equal $Q_\textrm{f}$ values in figures~\ref{atlas10} and~\ref{fig:fig13} are compared, the two sensor designs display no significant differences % it is evident that the PTI-design has smaller effect from 
at the strip implant edges. Thus, 
% Displayed in figure~\ref{fig:fig13b}, this is % solely 
% due to 
the doubled strip implant width, while maintaining equal interstrip gap size, of the Ioffe Institute-design sensor that % slows the growth of the 
results in lower average electric fields % peaks at the implant edges as a function of voltage.
in the area of the strip implant for the given biasing voltages, presented in figure~\ref{fig:fig13b}, has negligible effect on charge collection at the implant edge. Also the amplitudes of negative signals in the two figures match closely. % Thus, the experimentally observed about 10\% difference between the two sensor designs is % not verified by simulation.
% reproduced by the simulation only when ATLAS-design is considered with higher value of $Q_\textrm{f}$ to Ioffe Institute-design sensor.

% Overall 
As the figures~\ref{atlasM} and~\ref{fig:fig13} demonstrate, the simulated red laser scans with diameter matching the experimental set-up reproduce reasonably well the measured charge collection behavior of the 
% The observation of a negative charge signal in the case of red laser illumination of the p$^+$ strips is similar in both 
ATLAS- and Ioffe Institute-design strip detectors. % \cite{Eremin2003,Verbit2005}. % For deeper understanding of the observed dependence of the collected charge on the interface states $Q_\textrm{f}$ the scans were next repeated with 1 $\mu\textrm{m}$ laser to achieve results with higher resolution. % The important advantage of the PTI design is a window in metallization. % In this case the signal corresponding to position of the laser spot within the p$^+$ implant area that is free from metal is close to 100\%. That allows quantitative estimation of the signal changes in the laser scan.
%%%%%%%%%%%%%%%%%%%%%%%%%%%
% The charge collected in semiconductor detectors is the charge on the readout electrode at the end of the induction processes which exist while electrons and holes generated by the radiation drift towards the electrodes. In the case of signal read out from the p$^+$-contact, which collects holes, the collected charge signal is positive. This regularity is confirmed for the red laser illumination at the n$^+$-contact. In this case, the holes are collected by p$^+$-strips and induce a positive signal that fades to zero when the carrier injection point moves towards neighbouring strips \cite{Eremin2003}. % , as shown in Figs. 5a and b.
%
% Nevertheless, generating pairs on the strip side gives a result different to the expected one. %, as shown in Figs. 4a and b. The charge collected at the p$^+$-contact is positive when the charges are generated in the area near the readout strip but, as the injection point approaches the adjacent strip, the charge signals become negative instead of fading to zero. This can also be seen with the current pulses shown in figure~\ref{S13}, that present first positive polarity (charge injection closer to the readout strip) and then a negative polarity (charge injection closer to the strip adjacent to the readout strip), where bipolar pulses would be expected \cite{Radeka1988}. 
%
%
%%%%%%%%%%%%%%%%%%%%%%%%%%%%%%%%%%%%%%%%%%%%%%%%%%%%%%%%%%%%%%%%%%%%%%%%%%%%%%%%%
% \subsection{Interstrip scans with 1 $\mu\textrm{m}$ diameter red laser}
\subsection{Interstrip scans with reduced red laser diameter}
\label{1um}
For higher resolution study of the dependence of the collected charge on the interface states $Q_\textrm{f}$, % and the source of the reversed polarity or reduced signals, 
% For deeper understanding of the dependence of the collected charge on the interface states $Q_\textrm{f}$ and the source of the reversed polarity or reduced signals, 
observed in section~\ref{10um}, the scans were next repeated with reduced laser diameters between 1--3 $\mu\textrm{m}$, with 670 nm wavelength. % laser to achieve results with higher resolution. % Due to the identical behavior on both sides of the centermost strip seen in figures~\ref{fig:fig11} and~\ref{fig:fig13}, the following charge collection scans were focused between leftmost and center strips of the sensor structure.
Initial simulations with 1 $\mu\textrm{m}$ diameter laser proved to be problematic when laser position was at the collecting strip either at the strip-side of an implant bare from metallization, as in figures~\ref{S9} and~\ref{S9_2}, or at the backplane as in figure~\ref{fig:fig17b}. The complex calculations of the doping dependent SRH recombination in the region of high doping gradients resulted in unstable charge collection at the strip implant that was not redeemed by significantly increased mesh density in the modelled structure. This approach also increased the CPU time of each simulated charge injection point exponentially (e.g. mesh density increase by a factor of 3.5 leads to over 20 times longer simulation time), which is undesirable when the simulation study includes several hundred data points. However, increase of laser diameter from 1 to 3 $\mu\textrm{m}$ stabilized the simulated charge collection at high doping gradient regions considerably, while still providing substantially increased resolution to the charge injections of section~\ref{10um}. Thus, in aforementioned cases the following charge collection results were achieved by 3 $\mu\textrm{m}$ diameter laser illuminations.

\begin{figure}[tbp]
     \centering
%     \subfloat[Illumination with 1 $\mu\textrm{m}$ diameter red laser.]{\includegraphics[width=.65\textwidth]{Figures/Slide7.png}\label{Atlas1um}}\\
%     \subfloat[Zoom of (\ref{Atlas1um}).]{\includegraphics[width=.65\textwidth]{Figures/Slide7_zoom.png}\label{AtlasZoom}}
%     \subfloat[Illumination with 1 $\mu\textrm{m}$ diameter red laser.]{\includegraphics[width=.65\textwidth]{Figures/ATLAS_1um2.png}\label{Atlas1um}}\\
     \includegraphics[width=.65\textwidth]{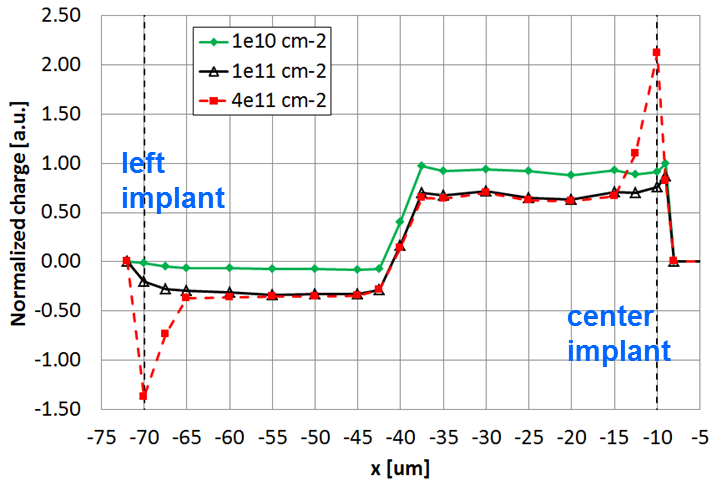} % \label{Atlas1um}\\
%     \subfloat[Zoom of (\ref{Atlas1um}).]{\includegraphics[width=.65\textwidth]{Figures/ATLASzoom.png}\label{AtlasZoom}}
    \caption{Corresponding red laser scan to figure~\ref{atlas10} with laser diameter decreased to 1 $\mu\textrm{m}$. Due to identical charge collection on both sides of the centermost strip, only left and center strips are plotted. % in (\ref{Atlas1um}) while (\ref{AtlasZoom}) provides a clearer view to differences in charge collection between varying values of $Q_\textrm{f}$ in the interstrip region. 
Charges are normalized to the maximum collection at $Q_\textrm{f}=1\times10^{10}$ $\textrm{cm}^{-2}$.}
\label{fig:fig12}
\end{figure}

\begin{figure}[tbp]
     \centering
%     \subfloat[Illumination with 1 $\mu\textrm{m}$ diameter red laser.]{\includegraphics[width=.65\textwidth]{Figures/Slide9.png}\label{S9}}\\
%     \subfloat[Al removed.]{\includegraphics[width=.65\textwidth]{Figures/CC_impl.png}\label{S9_2}}
     \subfloat[Illumination with 3 $\mu\textrm{m}$ diameter red laser.]{\includegraphics[width=.65\textwidth]{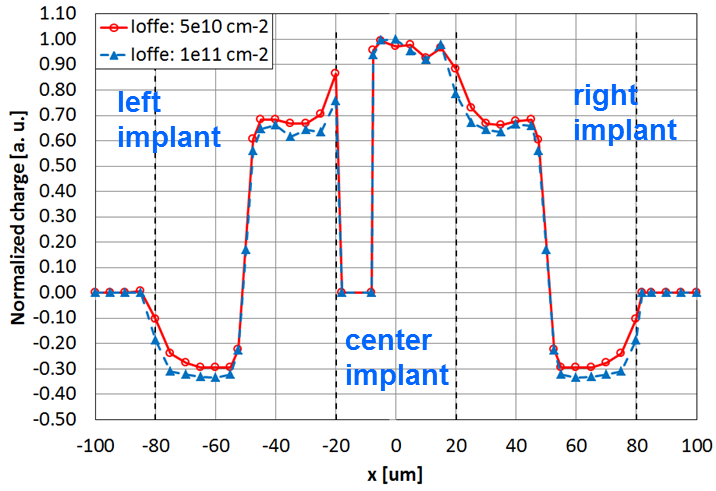}\label{S9}}\\
     \subfloat[Al removed.]{\includegraphics[width=.65\textwidth]{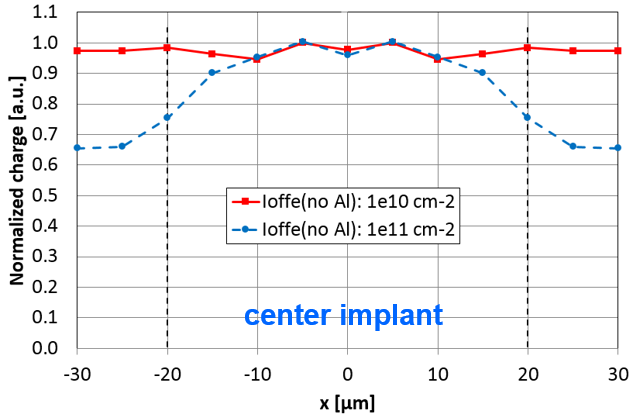}\label{S9_2}}
    \caption{(\ref{S9}) Corresponding red laser scan to figure~\ref{fig:fig13} with laser diameter decreased to 3 $\mu\textrm{m}$. (\ref{S9_2}) Red laser scan across the centermost strip implant when the strip metallization has been completely removed. Collected charges are normalized to the maximum collection at the lower value of $Q_\textrm{f}$ in each figure.}
\label{fig:fig15}
\end{figure}

The scan results for modified ATLAS-design strip detector in figure~\ref{fig:fig12} % with identical $Q_\textrm{f}$ steps to figure~\ref{fig:fig11}, 
are now significantly different to figure~\ref{atlas10} and reveal new features of the charge collection behavior between strips. As displayed in figure~\ref{fig:fig12} the charge collection of the highest $Q_\textrm{f}$ value is amplified substantially at the implant edges due to small spatial spread of the laser that results in the charge injection to be focused at the position of the electric field maxima.  

Also in figure~\ref{fig:fig12} it can be seen that the polarity reversal of the signal at midgap takes place abrubtly, within 5 $\mu\textrm{m}$. % The relative differences in % positive 
% collected charges, 
When the laser spot is above the oxide %, between % varying 
the two highest values of $Q_\textrm{f}$ % values 
are now % significantly smaller, the two highest values 
collecting almost equal amount of charge before % reaching the % center implant's 
the edges of the interstrip gap. Finally, it can be observed from figure~\ref{fig:fig12} that the effect of reversed polarity signals for laser positions beyond midgap from the readout strip is still present also for the lowest value of $Q_\textrm{f}$.

The laser scan results, with decreased laser diameter, % and increased number of laser positions, 
of the Ioffe Institute-design strip sensor in figure~\ref{fig:fig15} also display new features in charge collection behavior with respect to figure~\ref{fig:fig13}. The negative polarity signals in figure~\ref{S9} between % left 
adjacent and centermost strips at around $x=\lvert70\rvert~\mu\textrm{m}$ are now incresed by about 17\% and 12.5\% for $Q_\textrm{f}$ values $5\times10^{10}$ $\textrm{cm}^{-2}$ and $1\times10^{11}$ $\textrm{cm}^{-2}$, respectively. The charge collection at the lower value of $Q_\textrm{f}$ now reproduces closely the amplitudes of the measured reverse polarity signals in figure~\ref{fig:fig13}. As for the modified ATLAS-design, the polarity reversal at midgap is significantly more abrupt than in figure~\ref{fig:fig13}, taking place within 5 $\mu\textrm{m}$.

The amplitudes of the positive signals collected in the area of the oxide, with laser distance from the readout strip less than midgap, are also considerably lower, moving from about 70\% to 65\% with increased values of $Q_\textrm{f}$. % Notably 
For both values of $Q_\textrm{f}$, the positive signal collected at oxide reaches % its 
a local maximum close to midgap, immediately before polarity reversal, due to the high potential gradient close to the interstrip midgap potential maximum that increases the carrier drift velocities.

The charge collection plots in figure~\ref{S9} at the centermost strip have a % descending shape from the maximum value in the middle towards 
$Q_\textrm{f}$ dependence when the laser spot is getting closer to the right edge of the implant, free of metallization. To further investigate the $Q_\textrm{f}$ dependence when the laser spot is over the strip implant, %if this is an anomaly caused by the asymmetric positioning of the strip metallization, 
the scans at the readout strip were repeated without any metallization. The results in figure~\ref{S9_2} show that the charge collection of lower $Q_\textrm{f}$ fluctuates within 5\% with no descending pattern towards and beyond % behavior at the implant is now symmetrical, highest charges are collected at a longest distance to 
the implant edge. However, the curve for higher $Q_\textrm{f}$ has a descending shape towards the implant edge with about 25\% lower charge collected at the edge than % the charge collected at 
for lower value of $Q_\textrm{f}$. The charge collection of the two $Q_\textrm{f}$ values essentially converges within 10 $\mu\textrm{m}$ from the implant edge towards its center. Thus, % for 1 $\mu\textrm{m}$ red laser diameter 
the differences in $Q_\textrm{f}$ at the interstrip gap also affect the charge collection of a charge injection done over the implant (with strip metallization removed) up to few microns from the implant edge. 
%
% \begin{figure}[tbp] 
% \centering
% \includegraphics[width=.65\textwidth]{Figures/Slide8_Ex.png}
% \caption{Electric field distributions parallel to the detector surface at 0.8 $\mu\textrm{m}$ depth for the two detector designs. Voltages used for charge collection in both detectors are included, as well as $V=300$ V for the PTI detector for clearer comparison with the ATLAS detector.}
% \label{fig:fig14}
% \end{figure}
%
%%%%%%%%%%%%%%
\subsubsection{Critical interface charge density}
\label{CriticalQf}
The strong dependence of the reversed polarity signals, collected beyond midgap from the readout strip, on Si-SiO$_2$ interface charge density $Q_\textrm{f}$ is seen % especially 
clearly in the plots of sections~\ref{10um} and~\ref{1um}. This poses a question, is it possible to find a threshold level of $Q_\textrm{f}$ beyond which the effect of negative polarity signals vanishes. 

\begin{figure}[tbp] 
\centering
\includegraphics[width=.75\textwidth]{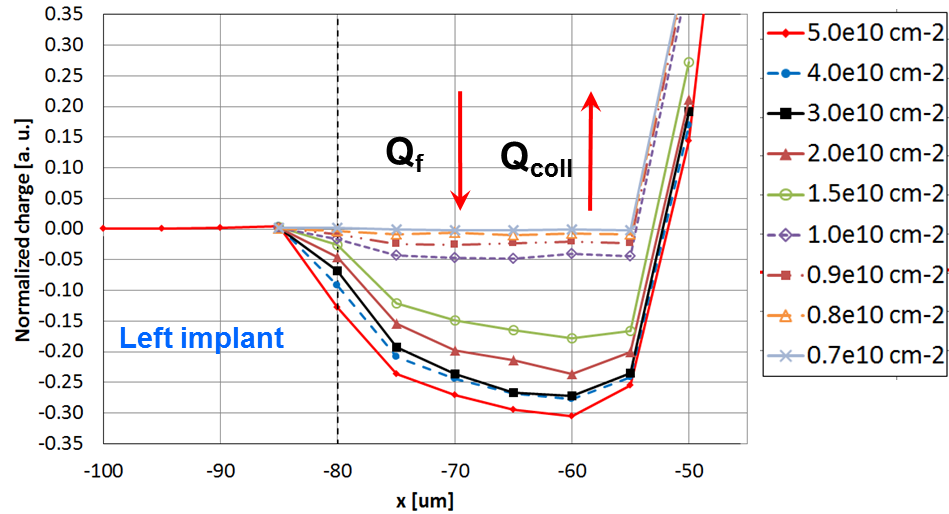}
\caption{Red laser scans for a threshold $Q_\textrm{f}$ where negative polarity signal vanishes. The detector parameters are as in figure~\ref{S9} with laser diameter of 1 $\mu\textrm{m}$. The collecting contact at -20 $\mu\textrm{m}$ (centermost strip) is not pictured. The red arrows indicate the dependence of the collected charge ($Q_\textrm{coll}$) on the interface charge density ($Q_\textrm{f}$).}
\label{fig:fig16}
\end{figure}

\begin{figure}[tbp] 
\centering
\includegraphics[width=.65\textwidth]{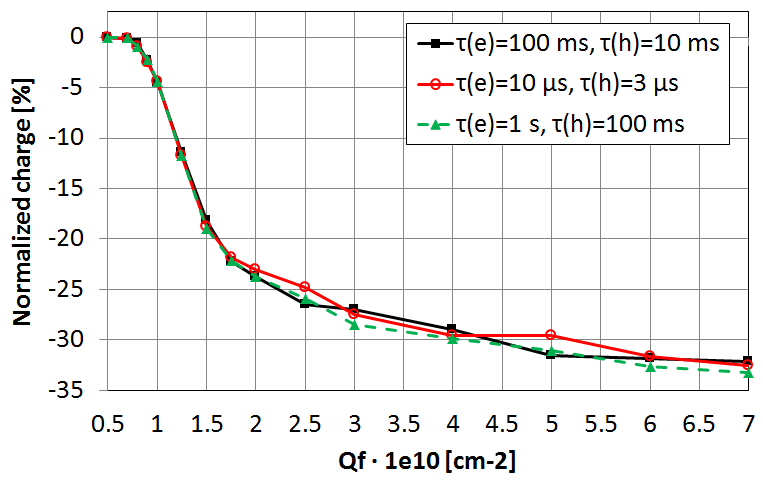}
\caption{Corresponding evolution of collected charge with $Q_\textrm{f}$ to figure~\ref{fig:fig16} with cut made at $x=-60~\mu\textrm{m}$. The negative polarity signal vanishes when $Q_\textrm{f}\leq0.8\times10^{10}~\textrm{cm}^{-2}$. Also included are curves for very low and very high, respectively, values of recombination lifetimes.% that are half and double of the initial lifetimes, respectively.
} 
\label{fig:fig17}
\end{figure}

The iteration of $Q_\textrm{f}$ values % from 
between $(0.5-7)\times10^{10}$ $\textrm{cm}^{-2}$, % to $0.5\times10^{10}$ $\textrm{cm}^{-2}$, 
for laser scans of a Ioffe Institute-design sensor in the region of the interstrip gap where negative polarity signals appear, is presented in figure~\ref{fig:fig16}. As can be seen, the negative charge collected at the centermost strip has a strong dependence on $Q_\textrm{f}$ and moves towards zero with decreased $Q_\textrm{f}$ values. When the collected charges at $x=-60~\mu\textrm{m}$, where their amplitude is at maximum in figure~\ref{fig:fig16}, are plotted as a function of $Q_\textrm{f}$ in figure~\ref{fig:fig17}, three regions % are observed. 
can be extracted. First, a region where negative signals settle to zero at $Q_\textrm{f}\leq0.8\times10^{10}~\textrm{cm}^{-2}$, then a strongly dynamic region of negative signal increase between $0.8\times10^{10}~\textrm{cm}^{-2}\textless{Q_\textrm{f}}\textless2.5\times10^{10}~\textrm{cm}^{-2}$, and ultimately a region with significantly reduced increase of the negative signal (about 23\% of the % increase at the 
dynamic region) for higher values of $Q_\textrm{f}$. 

When the % $Q_\textrm{f}$ scan is repeated for the 
recombination lifetimes % that are factor of two shorter and longer, respectively, to 
are varied from the initial % carrier lifetimes 
values (presented in the beginning of section~\ref{Results}) to very short and very long, respectively, the results seen in figure~\ref{fig:fig17} remain essentially the same. This can be expected since the ratio $t_\textrm{coll}/\tau_\textrm{tr}$ reflecting the level of trapping related charge losses (discussed at the start of this section) is still only $10^{-3}$ for the shortest recombination lifetime of electrons. % Since factor of two change in majority and minority carrier recombination lifetimes moves the simulated leakage current significantly out of range from experimentally observed values in non-irradiated p-on-n sensors but has negligible effect on the amplitudes of the collected negative polarity charges, this needs to be a fixed parameter.
% and ultimately a region close to a plateau for higher values of $Q_\textrm{f}$.

Thus, the negative polarity signals display an exclusive % strong
dependence on Si-SiO$_2$ interface charge states for a short range charge injection from the stripside at a given voltage and bulk doping. % and recombination lifetimes. 
The interstrip resistances remain at several tens of G$\Omega$ throughout the studied $Q_\textrm{f}$ range, which is in vicinity of experimentally observed values \cite{Bhardwaj2014}. Optimization of $Q_\textrm{f}$ does not then compromise the strip isolation, which in segmented p-on-n sensors is not a critical parameter since positive charges at the Si-SiO$_2$ interface do not attract holes (that could potentially open a conductive channel between the strips) to the accumulation layer.    
%%%%%%%%%%%%%%
\subsection{Interstrip scans from the backplane}
\label{Backplane}
\begin{figure}[tbp] 
\centering
\includegraphics[width=.65\textwidth]{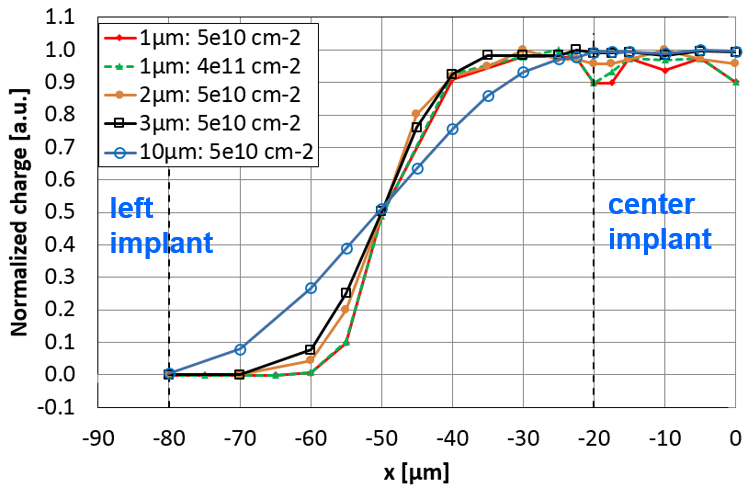}
\caption{Charges collected at the centermost strip of the Ioffe Institute strip detector % presented in figure~\ref{PTI} 
after a red laser scan from the backplane (n$^+$-side) % across three strips 
for two values of interface charge density $Q_\textrm{f}$. The charge collection was done at $V=400$ V while the laser diameter was varied from 1 to 10 $\mu\textrm{m}$. The strip implant edges are marked by dashed black lines. Collected charges are normalized separately for the two laser diameters to enable comparison.} %to the maximum collection at the lower value of $Q_\textrm{f}$ in each figure.}
\label{fig:fig17b}
\end{figure}

The red laser scans of the unsegmented n$^+$-side of an Ioffe Institute-design strip sensor were enabled by removing the metallization from the backplane of the modelled structure. The absence of the reversed polarity signals observed in earlier measurements \cite{Eremin2003} is reproduced by the simulation, as seen in figure~\ref{fig:fig17b}. No dependence on $Q_\textrm{f}$ is observed when the laser is moving below interstrip gap. The increase of charge sharing between the two strips in figure~\ref{fig:fig17b} as a function of laser diameter is also obvious.

The stability of the % collected charges 
charge collection simulation at the collecting strip also increases with laser diameter, as discussed in the beginning of section~\ref{1um}. By doubling the diameter from 1 $\mu\textrm{m}$ the charge collection dip at the edge of the collecting strip moves from 10\% to 4\%, while for 3 $\mu\textrm{m}$ diameter and above no fluctuations in the region of the collecting strip implant are observed and the curve shapes converge with measured data in Ref. \cite{Eremin2003}. 
%%%%%%%%%%%%%%%
% However, when approaching the readout strip at the center of the modelled sensor structure, the 1 $\mu\textrm{m}$ diameter laser % reveals new 
% displays features in the transient signals induced from the hole drift (as opposed to the electron drift from front surface injection), not seen in the measurement. The highest charge is now collected when the laser spot is below the oxide, some 5 $\mu\textrm{m}$ from the readout strip. Then a 10\% drop in the collected charge is observed at the implant edge and at the implant center. %, while the intermediate charges between the two local minima rise to about 3\% below the maximum collected charge. % Also a weak dependence on $Q_\textrm{f}$ is observed when the laser is moving below this region.
%
% When the backplane scan is repeated with a 10 $\mu\textrm{m}$ diameter laser that matches the experimental diameter, shown in figure~\ref{fig:fig17b}, no drop at the implant edge is observed and the measured behavior is reproduced closely. As in measurement \cite{Eremin2003}, when the 10 $\mu\textrm{m}$ diameter laser is moved across the interstrip gap the charges collected at the centermost strip never go to zero. This is opposite to the charge collection results of 1 $\mu\textrm{m}$ diameter laser, as can be seen from figure~\ref{fig:fig17b}. Thus, the spatial spread of the red laser is in important role in the shaping of charge collection distribution between and across the strips.
%%%%%%%%%%%%%%%

\begin{figure}[tbp] 
\centering
\includegraphics[width=.65\textwidth]{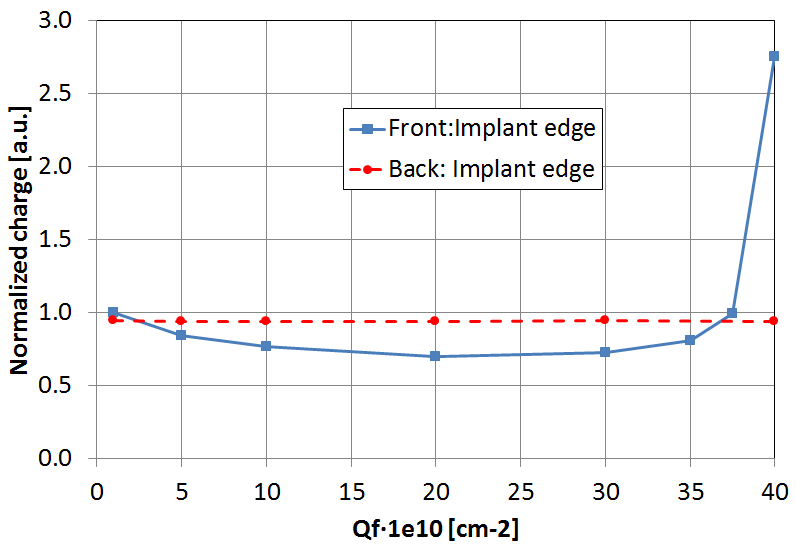}
\caption{Charges collected after red laser illumination both from strip-side and backplane at the implant edge of the collecting contact of the Ioffe Institute strip detector. The collected charges are normalized to the charge collected at the lowest value ($1\times10^{10}~\textrm{cm}^{-2}$) of interface charge density $Q_\textrm{f}$ from strip-side charge injection. The charge collection was done at $V=400$ V while the laser diameter was 1 $\mu\textrm{m}$.}
\label{fig:fig17c}
\end{figure}

To further investigate the dependencies of the charge collection behavior after % 1 $\mu\textrm{m}$ diameter 
red laser backplane illumination in the vicinity of the readout strip, the laser position was fixed to the implant edge ($x=-20~\mu\textrm{m}$ in figure~\ref{fig:fig17b}) and charge collection was monitored as a function of $Q_\textrm{f}$, as shown in figure~\ref{fig:fig17c}. As can be seen, for comparison the charge injection was also repeated from the strip-side. The $Q_\textrm{f}$ range was set wide enough to include % values below and above the 
experimentally observed values $Q_\textrm{f}=0.5-2\times10^{11}~\textrm{cm}^{-2}$ \cite{Robinson2002,Verbit2005}.

When the laser is at the strip-side in figure~\ref{fig:fig17c}, a clear charge multiplication (CM) effect is seen at $Q_\textrm{f}\textgreater3\times10^{11}~\textrm{cm}^{-2}$. 
At values below this, increased $Q_\textrm{f}$ results in significantly decreased positive collected charges due to increasing negative contribution from electron drift along the surface of the interstrip gap.
% At lower values of $Q_\textrm{f}$ an increasing negative contribution from surface drift with increased $Q_\textrm{f}$ results in significantly decreased collected positive charges. 
The increase of $Q_\textrm{f}$ increases also the electric field peaks at the implant edges. After the threshold value mentioned above this leads to CM of % the signal 
charge carriers that with further increase of $Q_\textrm{f}$ fully compensate for the decreased collected charge. %due to negative contribution from surface drift. 
However, when charge injection is done from backplane, the collected charge is completely independent from $Q_\textrm{f}$ and no CM is seen. Thus, when no physical processes account for the charge collection fluctuations of the 1 $\mu\textrm{m}$ red laser curves in figure~\ref{fig:fig17b} and since these vanish at increased laser diameter charge injections, the behavior can be deemed as simulators sensitivity to charge injections with very small spatial spread in the vicinity of high doping gradients, as discussed in the beginning of section~\ref{1um}.
% Thus, effects from these parameters cannot explain the 10\% drop in the collected charge at the implant edge in figure~\ref{fig:fig17b}.
% To decide whether or not the simulated charge collection behavior after 1 $\mu\textrm{m}$ diameter laser backplane illumination in the vicinity of the readout strip is a 
% The charge collection behavior from backplane injection in the vicinity of the readout strip gives an indication that not only $Q_\textrm{f}$ accounts for the charge losses seen in a strip sensor after short range charge injection.
%
\subsection{Transient signals}
\label{Trans}
To investigate the time evolution of the transient signals induced by the red % and IR 
laser, two laser spot positions were considered, 5 $\mu\textrm{m}$ from the edge of the readout strip and 5 $\mu\textrm{m}$ from the edge of its adjacent strip ($\lvert x\rvert=25~\mu\textrm{m}, 75~\mu\textrm{m}$ in figure~\ref{S9}, respectively). These positions enable the study of transient currents leading to both positive and negative collected charges in the case of short range charge injection from the strip-side. Since the current response is
\begin{equation}%\label{eq.1}
% i(t)=\frac{\textrm{d}}{\textrm{d}t}Q_{\textrm{coll}}(t),
i(t)=\frac{\textrm{d}}{\textrm{d}t}Q(t),
\end{equation}
the features of the collected charge distribution as a function of position, % described above 
presented in the previous sections, can also be seen in the transient current plots. % current responses 
% The charge collected in semiconductor detectors is the charge on the readout electrode at the end of the induction processes which exist while electrons and holes generated by the radiation drift towards the electrodes. In the case of signal read out from the p$^+$-contact, which collects holes, the collected charge signal is positive. This regularity is confirmed for the red laser illumination at the n$^+$-contact. In this case, the holes are collected by p$^+$-strips and induce a positive signal that fades to zero when the carrier injection point moves towards neighbouring strips \cite{Eremin2003}. % , as shown in Figs. 5a and b.
%
% Nevertheless, generating pairs on the strip side gives a result different to the expected one. %, as shown in Figs. 4a and b. 
% The charge collected at the p$^+$-contact is positive when the charges are generated in the area near the readout strip but, as the injection point approaches the adjacent strip, the charge signals become negative instead of fading to zero. This can also be seen with 

\begin{figure}[tbp]
     \centering
     \includegraphics[width=.6\textwidth]{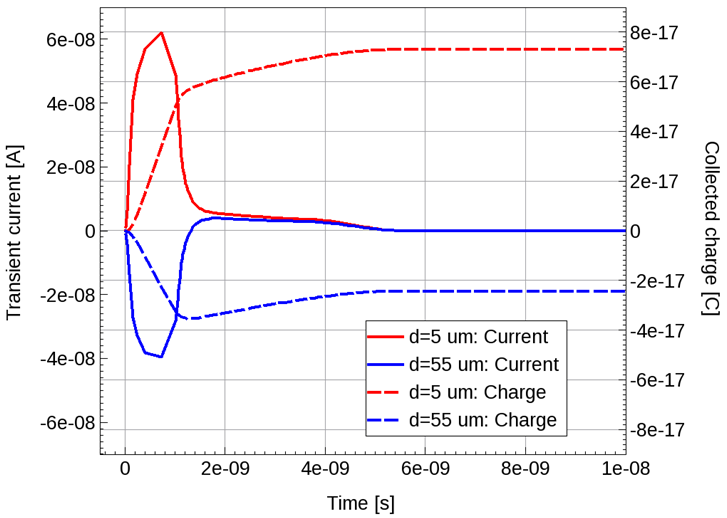}\\ %\label{S13}\\
%     \subfloat[Infra-red laser illumination.]{\includegraphics[width=.6\textwidth]{Figures/Slide13_IR.png}\label{S13_2}}
    \caption{Transient current signals and collected charges after % (\ref{S13}) 
red laser % and (\ref{S13_2}) infra-red laser 
illuminations. Two laser positions are considered, 5 $\mu\textrm{m}$ and 55 $\mu\textrm{m}$ from the collecting contact when the interstrip gap is 60 $\mu\textrm{m}$. The interface charge density % for ($\ref{S13}$) 
was $Q_\textrm{f}=1\times10^{11}~\textrm{cm}^{-2}$.} % and for (\ref{S13_2}) $Q_\textrm{f}=5\times10^{10}~\textrm{cm}^{-2}$.}
\label{fig:fig19}
\end{figure}

The current pulses in a red laser illuminated Ioffe Institute-design strip sensor at $V=400~\textrm{V}$ in figure~\ref{fig:fig19} show that %that present first positive polarity (charge injection closer to the readout strip) and then a negative polarity (charge injection closer to the strip adjacent to the readout strip), where bipolar pulses would be expected \cite{Radeka1988}. 
%
%The features of the collected charge distribution vs. position described above are also seen in the current responses %. Note here that the current response is
% \begin{equation}%\label{eq.1}
% i(t)=\frac{\textrm{d}}{\textrm{d}t}Q_{\textrm{coll}}(t).
% \end{equation}
% For red laser injection on the p$^+$-side in the area close to the Al metallization the positive charge collected at the centermost strip corresponds to a positive polarity current pulse with a gradual decay. In the interstrip gap the current response decreases in amplitude and then inverts polarity at the midgap. The amplitude of the negative pulse then increases, reaching a maximum at the border of the p$^+$-implant of the adjacent strip. 
the duration of the negative current response is identical to the width of the positive current response. Also distinct is the smaller amplitude of the negative current pulse when $t\textless1.3~\textrm{ns}$. Both pulses induced by the electron drift have identical low-amplitude positive tail that goes to zero after 5 ns. As can be seen from the time integrals of the transient currents, i.e. induced charges, this results in the reduction of the negative charge until the charge collection is complete.

\begin{figure}[tbp] 
\centering
 \subfloat[Electron current density at $t=0.75~\textrm{ns}$.]{\includegraphics[width=.5\textwidth]{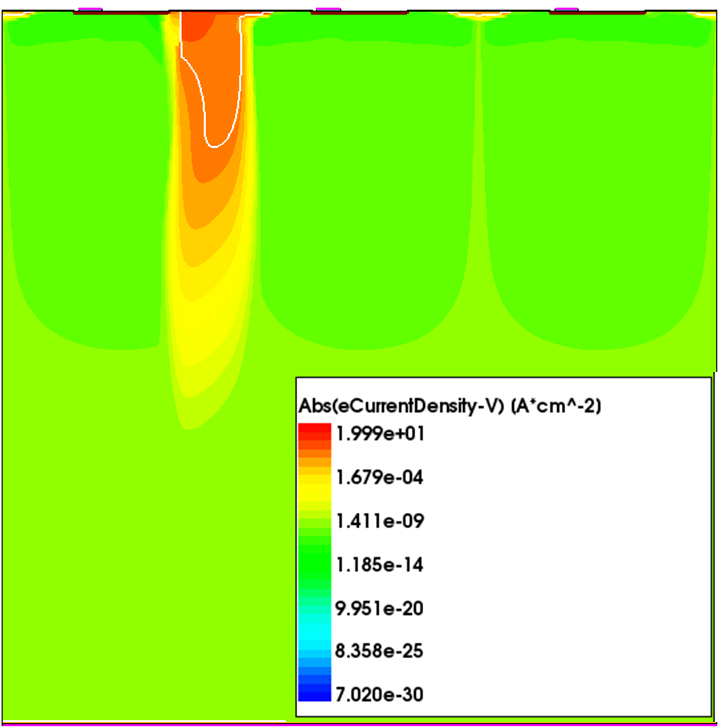}\label{ec075}}
 \subfloat[Electron current density at $t=1.0~\textrm{ns}$.]{\includegraphics[width=.5\textwidth]{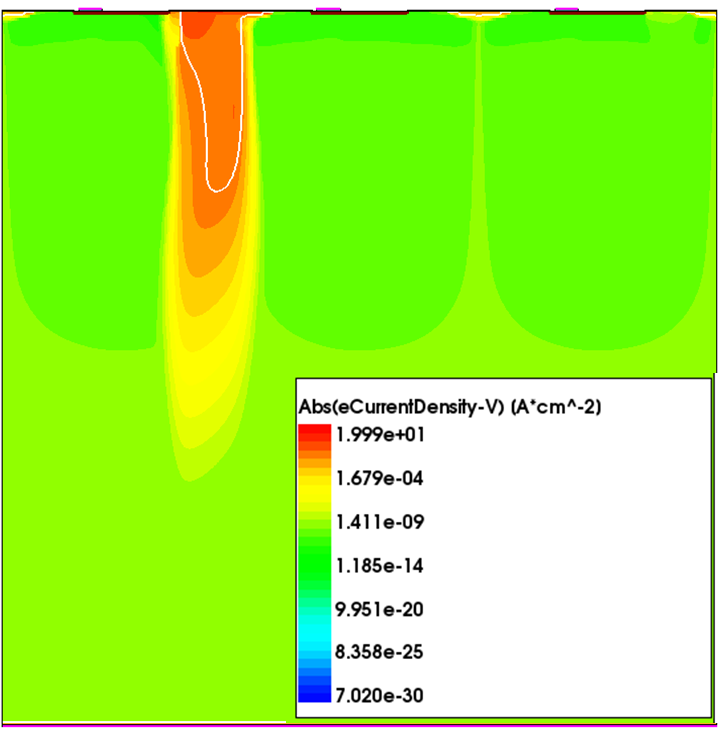}\label{ec1}}\\
 \subfloat[Electron current density at $t=1.5~\textrm{ns}$.]{\includegraphics[width=.5\textwidth]{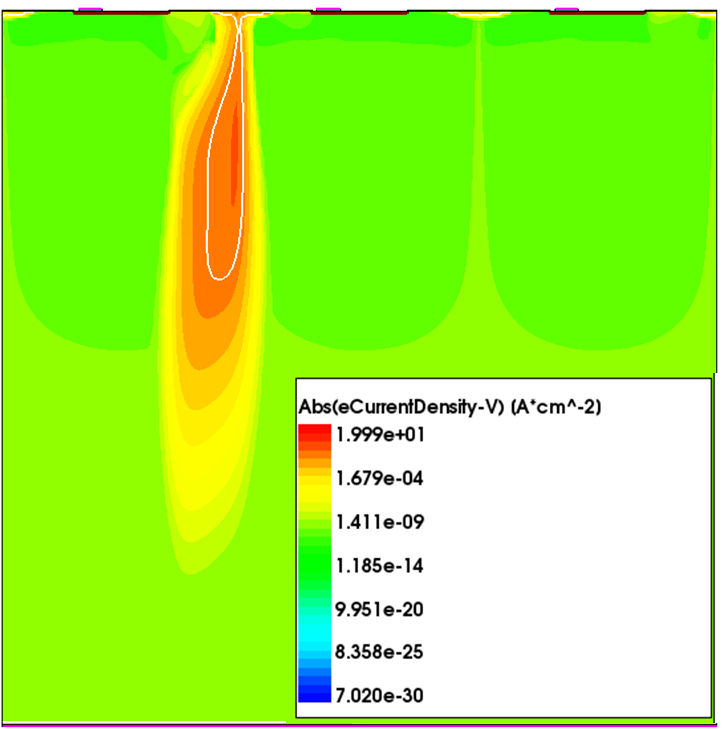}\label{ec15}}
 \subfloat[Electron current density at $t=2.0~\textrm{ns}$.]{\includegraphics[width=.5\textwidth]{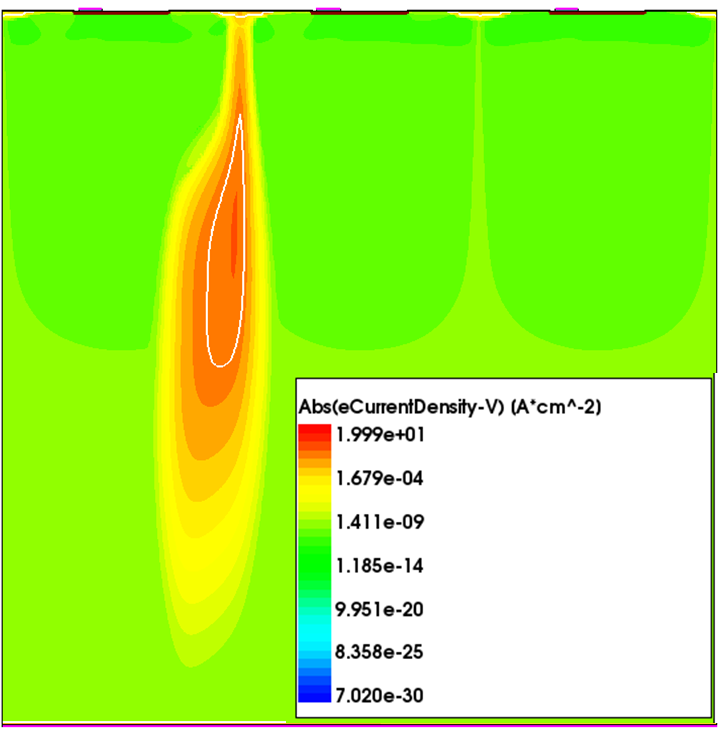}\label{ec2}}
\caption{% Charges collected at the centermost strip of the ATLAS strip detector presented in figure~\ref{ATLAS1} after a red laser scan across three strips for (a) measured and simulated sensor 
The electron current densities corresponding to $d=55~\mu\textrm{m}$ in figure~\ref{fig:fig19} for the time interval 0.75--2 ns.
% Charges collected at the centermost strip after a red laser scan across the three strips of (a) measured and simulated ATLAS strip detector presented in figure~\ref{ATLAS1} and for (b) varied values of interface charge density $Q_\textrm{f}$ in a simulated strip sensor structure modified from ATLAS-design presented in figure~\ref{ATLAS}. % that is in units of cm$^{-2}$. The bias voltages during the charge collection were (a) 350 V and (b) 300 V while the laser diameter was 10 $\mu\textrm{m}$. The strip implant edges are marked by dashed black lines. The simulated charges in (b) are normalized to the maximum charge collected at $Q_\textrm{f}=1\times10^{10}~\textrm{cm}^{-2}$. The measured curve in (a) is from Ref. \cite{Eremin2003}.
}
\label{eCD}
\end{figure}

Since the polarity reversal of the transient current signal for $d=55~\mu\textrm{m}$ in figure~\ref{fig:fig19} takes place at $t=1.3~\textrm{ns}$, the time evolution of electron current densities before and after this point were plotted in figure~\ref{eCD} % to illustrate the carrier movements responsible for the transient signals.
as being the most revealing interval for understanding the formation of the transient signals. As the charge injection point in figure~\ref{eCD} is $d=55~\mu\textrm{m}$ from the centermost strip ($5~\mu\textrm{m}$ from the leftmost strip) the % formation of both 
two transient signals in figure~\ref{fig:fig19} can be considered to be % studied 
simultaneously % by considering
induced at the two adjacent strips from this charge injection position.

Figures~\ref{ec075} and~\ref{ec1} show that before the polarity reversal of the negative transient signal in figure~\ref{fig:fig19} the highest electron current densities are seen between the leftmost strip closest to the charge injection point and the midgap between the leftmost and center strips of the 3-strip sensor structure. The induced negative transient signal at the centermost strip is then solely generated by the electron drift from the potential minimum of the leftmost strip to the local potential maximum at the midgap, while the positive transient signal seen at the leftmost strip is a composite of electron drift toward global and local potential maxima at the sensor backplane and midgap, respectively. The transient signal at the leftmost strip is then the result of positive signal induced from drift towards the sensor backplane reduced by opposite sign signal induced from the drift parallel to the sensor surface having equal amplitude to the transient signal seen at the centermost strip.

From figures~\ref{ec15} and~\ref{ec2} it can be seen that %show that before After 
the polarity reversal from negative to positive %, i.e. at $t\textgreater1.3~\textrm{ns}$ 
in figure~\ref{fig:fig19} is the result of completed electron drift parallel to the sensor surface to the local potential maximum at the midgap. After this, i.e. at $t\textgreater1.3~\textrm{ns}$, only electron current densities normal to the sensor surface are observed. Since these are generated essentially from the electron drift at equidistance from the leftmost and center strips, the induced transient signals at both strips have identical sign and amplitude. 
% The figure~\ref{S13_2} displays the transient currents and collected charges in the same sensor structure after IR laser illumination, that generates charge carriers uniformly throughout the sensor bulk thickness. Again the transient current signals generated at the two opposing laser spot positions are identical in duration that is now about 8.5 ns due to low mobility holes contributing to the signal while drifting through the sensor thickness to the readout strip. 
%
% The charge injection position further away from the readout strip results now in bipolar transient current signal that amounts in essentially zero collected charge, that can also be seen in figure~\ref{fig:fig18} for the same value of $Q_\textrm{f}$ at $x=-75~\mu\textrm{m}$ (55 $\mu\textrm{m}$ from the collecting strip).
%

\section{Discussion}
\label{Discussion}
To provide an interpretation of the observed results in this study, % two baselines need to be established. First, 
it is considered that when the readout contact is a cathode, as in p-on-n strip sensors, holes drifting towards it and electrons drifting away from it induce a positive transient current that accumulates a positive collected charge. % Second, when the readout contact is an anode, as in n-on-p strip sensors, holes drifting away from it and electrons towards it induce a positive transient current that accumulates a positive collected charge. Reversed directions in both cases result in reversed polarity charge signals.
Reversed directions result in reversed polarity charge signals.

\begin{figure}[tbp]
     \centering
%     \subfloat[Illumination with 1 $\mu\textrm{m}$ diameter red laser.]{\includegraphics[width=.65\textwidth]{Figures/Slide9.png}\label{S9}}\\
%     \subfloat[Al removed.]{\includegraphics[width=.65\textwidth]{Figures/CC_impl.png}\label{S9_2}}
     \subfloat[Potential distribution in the strip sensor modified from ATLAS-design.]{\includegraphics[width=.65\textwidth]{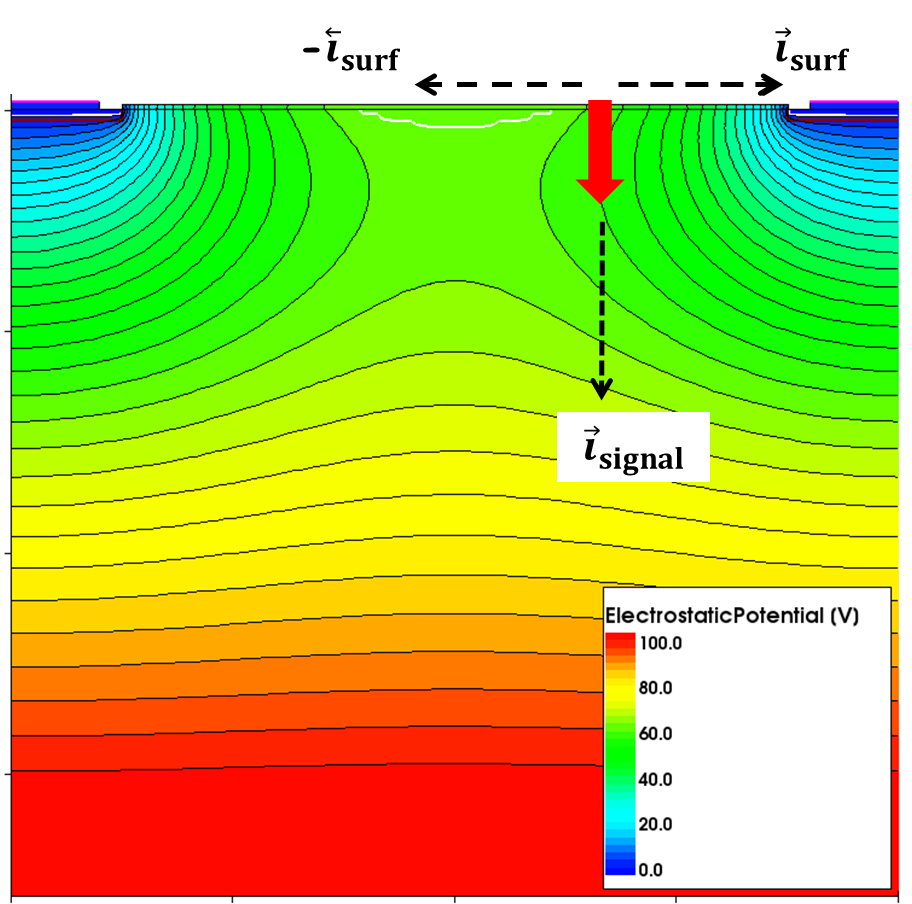}\label{Pot2}}\\
     \subfloat[Potential cut at midgap through sensor bulk.]{\includegraphics[width=.65\textwidth]{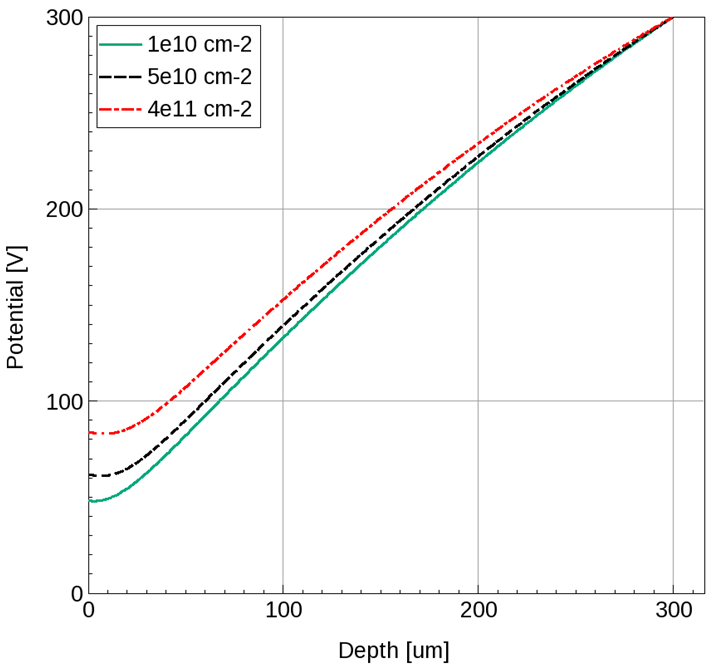}\label{Pot}}
    \caption{Electrostatic potential distribution at $V=300~\textrm{V}$ up to 70 $\mu\textrm{m}$ depth (\ref{Pot2}) and the corresponding cut through the sensor bulk (\ref{Pot}) at the midgap between strips for varying values of $Q_\textrm{f}$ in modified ATLAS strip sensor structure. (\ref{Pot2}) also depicts an example of a charge injection by a red laser (red arrow) through the interstrip oxide layer and the resulting components of transient current.} 
% (\ref{S9}) Corresponding red laser scan to figure~\ref{fig:fig13} with laser diameter decreased to 1 $\mu\textrm{m}$. (\ref{S9_2}) Red laser scan across the centermost strip implant when the strip metallization has been completely removed. Collected charges are normalized to the maximum collection at the lower value of $Q_\textrm{f}$ in each figure.}
\label{fig:fig20}
\end{figure}

% The explanation % of the induction delay is 
These statements are just consequences % can be based on 
of the Ramo's theorem \cite{Cavalleri1971,Gatti1982}, which
gives an accurate result for the induced current, that is the charge induction rate, as 
\begin{equation}%\label{eq.1}
%\frac{\textrm{d}Q}{\textrm{d}t}=\frac{\partial}{\partial\xi}\int_{A} E_{\textrm{q}}(d)\textrm{d}s\frac{\textrm{d}\xi}{\textrm{d}t}.
i=-qE^{*}v=\frac{\textrm{d}Q}{\textrm{d}t},
\end{equation}
%
% a product of 
where $q=\pm{e}$ is the carrier charge with $e$ as the elementary charge, $E^\ast$ is the weighting electric field and % the charge drift velocity 
$v$ is the charge drift velocity.
\begin{enumerate}
%%%%%%%%%%%%%%%%%%%%%%%%
\item The reduction of the laser diameter from the experimentally used value \cite{Eremin2003,Verbit2005} to 1--3 $\mu\textrm{m}$ in figures~\ref{fig:fig11} and~\ref{fig:fig12} for modified ATLAS-design as well as in figures~\ref{fig:fig13},~\ref{S9} and~\ref{fig:fig17b} for Ioffe Institute-design, % resulted in several distinct effects.  This can be seen as the result of the smaller spatial distribution of the charge carrier generation that 
lead to several times smaller averaging of the effects to the carrier transport from the electric field distribution and drift distances parallel to the sensor surface. This is seen in figures~\ref{fig:fig12} and~\ref{S9} % First, when the laser spot position was moved across the interstrip oxide, a 
in the decrease of positive collected charges and increase of the absolute value of the negative polarity charges when the laser spot position was moved across the interstrip oxide, as well as % For the same reason 
sharper polarity reversal at the midgap and higher sensitivity to the edges of the strip implants. 
\item The effects of smaller diameter laser spot also include % slight 
a visible reduction of the $Q_\textrm{f}$ dependence, i.e. the charges collected from laser illumination over the interstrip oxide have % smaller 
decreasing differences % between varied values of 
with increased $Q_\textrm{f}$. % Since increased $Q_\textrm{f}$ will result in higher electric field peaks at the strip implant edges in a p-on-n strip sensor\footnote{Higher value of the positive oxide charge density will attract more electrons to the accumulation layer at the Si-SiO$_2$ interface. The electrons are not collected by the cathode strip electrodes, leading to increased electric fields close to the strip edges.}, whether or not these areas are included in the charge generation region (by the size of the laser diameter) will influence the dependence of the collected charge on the value of $Q_\textrm{f}$. 
Evidence from figures~\ref{atlas10},~\ref{fig:fig12} and~\ref{fig:fig17} % seems to also 
indicates that for the smaller diameter charge injection there is an upper threshold value of $Q_\textrm{f}$ beyond which only limited increase in negative polarity signals is seen with increased % values of 
$Q_\textrm{f}$ in the lower field regions of the interstrip gap. 

\item % In figure~\ref{S9_2} the increased resolution of the laser displays % more clearly 
% in figure~\ref{S9_2} the % position 
Figure~\ref{S9_2} displays the $Q_\textrm{f}$ dependence of the collected charge when the laser spot is moved over the readout strip implant % of the readout strip 
close to its edge % on a Ioffe Institute-design strip sensor with % charge injection window in the 
when the strip metallization has been removed. The order of magnitude difference in $Q_\textrm{f}$ for the two curves % has practically no effect on 
results in about 25\% lower collected charge for the higher $Q_\textrm{f}$ at the edge of the strip implant while the $Q_\textrm{f}$ dependence essentially vanishes when the 3 $\mu\textrm{m}$ diameter laser spot is moved 10 $\mu\textrm{m}$ from the implant edge towards its center. 

% It % is important to stress 
% should be emphasized that the losses of the laser light in the p$^+$ implant do not exceed 10\% which is % much 
% significantly less than the drop of about 30\% of CCE for the two higher $Q_\textrm{f}$ values in figure~\ref{fig:fig15} when the laser spot is moved to the edge of the collecting strip implant and over the interstrip oxide. 
%
\item The interpretation of earlier experimental observations concluded % As a result, 
that the oxide layer with the attractive electric field may be considered as a "sink" for electrons that are moving close to the surface %. Obviously the electrons, which are accumulated in the layer, will reduce the effect of attraction for the drifting electrons, however 
with the probability to reach the 
surface and be trapped in the interstrip gap significantly higher than for those, which are generated under the p$^+$ implant \cite{Verbit2005}. Since substantial reduction of the collected charge is % now 
observed at % the location of the p$^+$ implant with insignificant dependence on $Q_\textrm{f}$, % this interpretation needs to be complemented by looking at 
laser positions less than midgap distance from the collecting strip, the total transient current signal can be considered as a composite of the currents induced by the carrier drifts along the surface ($i_\textrm{surf}$) and orthogonal to the surface through the sensor thickness ($i_\textrm{signal}$) as 
\begin{equation}\label{eq.trans}
%\frac{\textrm{d}Q}{\textrm{d}t}=\frac{\partial}{\partial\xi}\int_{A} E_{\textrm{q}}(d)\textrm{d}s\frac{\textrm{d}\xi}{\textrm{d}t}.
i_\textrm{tot}=i_\textrm{signal}+i_\textrm{surf}.
\end{equation}
These are illustrated in figure~\ref{Pot2}. With the statements from the first paragraph of this section and equation (\ref{eq.trans}) the curves in figure~\ref{S9_2} can now be interpreted of having minimal contribution from the opposite sign $i_\textrm{surf}$ to $i_\textrm{tot}$ in the middle of the p$^+$ implant, while its share increases when the point of charge injection moves 
to the implant edge, i.e. when the electron drift along the surface of the interstrip gap changes from negligible to significant. 

If $i_\textrm{surf}$ is broken into two components with opposite velocity vectors, as in figure~\ref{Pot2}, % this in turn 
the increase in collected charges close to midgap in figures~\ref{fig:fig12} and~\ref{S9} can be interpreted to be the result of the components compensating each other % in the middle where both have equal 
with close to equal amplitudes. As a result, in the middle of the interstrip gap $i_\textrm{tot}$ has % contribution only 
decreased contribution from $i_\textrm{surf}$. When the position of charge injection moves closer to the implant edge of the collecting contact, the two opposite components of $i_\textrm{surf}$ become increasingly non-equal, leading to growing opposite sign contribution to $i_\textrm{tot}$.

% \item To understand the 10\% decrease in collected charge after red laser illumination from the sensor backplane at a position opposite to the implant edge of the collecting strip in figure~\ref{fig:fig17b}, the investigation presented in figure~\ref{fig:fig17c} revealed the effect to be $Q_\textrm{f}$ independent, i.e. the holes drifting through the sensor bulk are not interacting with the layer of accumulated electrons % accumulation layer electrons 
% in the interstrip gap. Thus, the positive $I_\textrm{signal}$ induced by the hole drift towards the cathode strip is then the sole component of $I_\textrm{tot}$. In addition, the datapoint at the implant edge does not constitute any descending or ascending trend in the charge collection curves and when the backplane scan is repeated with a 10 $\mu\textrm{m}$ diameter red laser the effect vanishes. Hence, the lower collected charge at the implant edge is not regarded as a physical effect but a simulation artifact as the few percent fluctuations in figure~\ref{S9_2}. 
%
\item The strip-side red laser illuminations % from the sensor backplane at a position opposite to 
at the implant edge of the collecting strip in figure~\ref{fig:fig17c} show strong dependence on $Q_\textrm{f}$. Due to the charge injection position, the only possible direction for $i_\textrm{surf}$ is away from the collecting strip. Thus, even though both $i_\textrm{surf}$ and $i_\textrm{signal}$ are now induced from the drift away from the collecting contact, $i_\textrm{surf}$ towards the adjacent strip and $i_\textrm{signal}$ towards backplane, they clearly induce opposite polarity signals. Hence, % the polarity of $I_\textrm{surf}$ is determined relative to $I_\textrm{signal}$, i.e. 
as in the above interpretation of compensation of the two components of $i_\textrm{surf}$, the equal sign charge carriers with different direction velocity vectors compensate each other. 

\item Following the model proposed in \cite{Verbit2005} the curvature of the electric field under SiO$_2$ in interstrip gap has a minimal potential for electrons at the interface and a saddle point with local potential maximum % electric field 
at certain distance in the bulk. Such a potential distribution can act as sink for the electrons. % For sure 
Undoubtedly the local potential maximum % (minimal potential difference between backplane and sensor surface) 
seen in figures~\ref{Pot2} and~\ref{Pot} works against the drift of the electron cloud to the interface, however, some fraction of the cloud can diffuse towards the interface via the potential minimum. Obviously the shape of the peak is important for the charge losses and the shape changes with $Q_\textrm{f}$, as can be seen from figure~\ref{Pot}.

% The earlier interpretation that the potential minimum at the silicon surface in the interstrip gap caused by the positive oxide charge % in p-on-n strip sensors %. The regions % (layers) with potential minima can be considered as "sinks" for the drifting electrons % in the drifting "cloud" \cite{Verbit2005}, 
The influence of sinks to the charge collection can also be concluded from figure~\ref{fig:fig19}. The transient current curve for $d=55~\mu\textrm{m}$ (5 $\mu\textrm{m}$ from the strip adjacent to the readout strip) displays first a higher amplitude negative polarity signal due to electron drift from the high electric field region, repulsed by the injected electrons, towards the readout strip. Then a low amplitude positive 'tail' is induced due to the drift of the electrons, not trapped at the % local 
potential minimum, away from the readout strip towards the sensor backplane. Identical 'tail' is observed for $d=5~\mu\textrm{m}$ transient curve in figure~\ref{fig:fig19}, showing that the accumulation layer electrons are again repulsed to the local potential maximum region providing the low amplitude, slow component to the positive transient signal. % Since the uncompensated negative component was also observed in p-spray isolated n-on-p strip sensor in figure~\ref{S15}, it is clear that also holes experience trapping produced imbalance between negative $I_\textrm{surf}$ and positive $I_\textrm{signal}$. 

% Following the model proposed in \cite{Verbit2005} the curvature of the electric field under SiO$_2$ in interstrip gap has a minimal potential for electrons at the interface and the saddle point with maximal electric field at certain distance in the bulk. Such a potential distribution can act as sink for the electrons. % For sure 
% Undoubtedly the potential maximum works against the drift of electron cloud to the interface, however, some fraction of the cloud can diffuse towards the interface via the  potential maximum. Obviously the shape of the peak is important for the charge losses and the shape changes with $Q_\textrm{f}$. \\ \\

\item As was shown in figures~\ref{fig:fig16} and~\ref{fig:fig17}, it is possible to find values of Si-SiO$_2$ interface charge density $Q_\textrm{f}$ within an order of magnitude of typical values seen in non-irradiated strip sensors ($(0.2-2)\times10^{11}~\textrm{cm}^{-2}$ \cite{Robinson2002,Verbit2005,Poehlsen2013,Moscatelli2016}) that can reduce negative $i_\textrm{surf}$ to an insignificant level in a p-on-n strip sensor. 
% In figure~\ref{fig:fig16} the critical Si-SiO$_2$ interface charge density was defined and 
In the particular cases of ATLAS and alternative Ioffe Institute topologies with different strip pitches and strip implant widths % of interstrip gap 
the values are in the range of $(0.8-1.5)\times10^{10}~\textrm{cm}^{-2}$.

\item The resulting observations lead to the conclusion, that a charge injection through % any higher level carrier density or concentration close to the surface 
Si-SiO$_2$ accumulation layer of a segmented p-on-n silicon sensor induces $i_\textrm{surf}$ of which sign is always opposite relative to $i_\textrm{signal}$.
% Its sign is determined by the direction of the dominating carrier drift parallel to the surface with respect to the collecting contact and the biasing potential of the contact. 
The level of influence from $i_\textrm{surf}$ to $i_\textrm{tot}$ is governed by the localization in the potential minimum originated from $Q_\textrm{f}$ at Si-SiO$_2$ interface that manifests as an
% trapping produced 
imbalance between the two opposing components of $i_\textrm{tot}$ induced by the drifts parallel and normal to the sensor surface. 
%
% In the absence of $i_\textrm{signal}$, i.e. at charge injection positions beyond midgap from the collecting strip, the reversed polarity of $i_\textrm{surf}$ is due to electron drift towards the collecting cathode of the p-on-n strip sensor.
\end{enumerate}

\section{Conclusions}
\label{Conclusions}
The performed simulations of the influence of Si-SiO$_2$ interface charge on the current response and charge collection efficiency reproduces all experimental features, i.e. negative fraction in the charge response and tens of percents reduction of CCE which were observed earlier \cite{Eremin2003,Verbit2005} in non-irradiated Si strip detectors. 
%
% To understand the nature and source of the reversed polarity charge signals observed in the earlier measurements of p-on-n strip detectors illuminated by a red laser with varying positions across strip pitch \cite{Eremin2003,Verbit2005}, the measurements were first reproduced by simulation using parameters close to the experimental set-up. 
When % satisfactory 
agreement between measurement and simulation was obvious % in figures~\ref{fig:fig11} and~\ref{fig:fig13}, 
the simulations could be extended to parameter values beyond the experimental study with justified confidence on the results. 
The study demonstrated that:
\begin{enumerate}
\item Variation of the interface charge and its influence on the detector response in the performed modelling gives clear indication that the features are originated from the interface charge and are defined by its density. 
\item There is a critical density of the interface charge which initiates the features and the value of critical density is not sensitive to the topology of the strip detector. 
\item The critical interstrip charge density was defined and in the particular cases of ATLAS and alternative Ioffe Institute topologies, with different strip pitches and strip implant widths, % of interstrip gap 
the values are in the range of $(0.8-1.5)\times10^{10}~\textrm{cm}^{-2}$. 
\item Simulation of the charge collection in the interstrip gap with the laser beam spot equal to that used in the experiment (10 $\mu\textrm{m}$ diameter) or % ten 
several times less showed that the negative response in the active strip changes sharply its value in the middle of the interstrip gap. This shows a direct link to the trajectory of the collected charge cloud. It is important to which strip the electric field line and the related trajectory of the cloud flow. % in.  
\end{enumerate}
The performed study is a physical base for explanation and prediction of the influence of radiation on the detector performance. Generation of the charge % in SiO$_2$-layer and 
at the Si-SiO$_2$ interface is a specific process sensitive to the properties of SiO$_2$ layer and its thickness. The results show the possibility for the irradiation effect stabilization and/or its minimization via optimization of interstrip gap passivation. 

The results provide grounds to extend the study 
% The investigation of 
to the experimentally observed absence of reversed polarity signals in infra-red laser (with minimum ionizing particle-like carrier generation) %, with a penetration depth of several hundred microns in silicon, 
illuminated strip sensors \cite{Eremin2003}, as well as the possible sensitivity of n-on-p strip sensors to the effect, which is a strict plan for the nearest future. 

\bibliography{mybibfile}

\begin{thebibliography}{10}
\expandafter\ifx\csname url\endcsname\relax
  \def\url#1{\texttt{#1}}\fi
\expandafter\ifx\csname urlprefix\endcsname\relax\def\urlprefix{URL }\fi
\expandafter\ifx\csname href\endcsname\relax
  \def\href#1#2{#2} \def\path#1{#1}\fi

\bibitem{Robinson2002}
D.~Robinson, et~al., {S}ilicon microstrip detectors for the {ATLAS SCT}, Nucl.
  Instr. $\&$ Meth. A 485 (2002) 84--88.
\newblock \href {http://dx.doi.org/10.1016/S0168-9002(02)00536-3}
  {\path{doi:10.1016/S0168-9002(02)00536-3}}.

\bibitem{Andricek2000}
L.~Andricek, et~al., {D}esign and test of radiation hard p$^+$n silicon strip
  detectors for the {ATLAS SCT}, Nucl. Instr. $\&$ Meth. A 439 (2000) 427--441.
\newblock \href {http://dx.doi.org/10.1016/S0168-9002(99)00901-8}
  {\path{doi:10.1016/S0168-9002(99)00901-8}}.

\bibitem{Allport2000}
P.~P. Allport, et~al., {A} comparison of the performance of irradiated p-in-n
  and n-in-n silicon microstrip detectors read out with fast binary
  electronics, Nucl. Instr. $\&$ Meth. A 450 (2000) 297--306.
\newblock \href {http://dx.doi.org/10.1016/S0168-9002(00)00259-X}
  {\path{doi:10.1016/S0168-9002(00)00259-X}}.

\bibitem{Kraner1983}
H.~W. Kraner, et~al., {C}harge {C}ollection in {S}ilicon {S}trip {D}etectors,
  IEEE Trans. Nucl. Sci 30 (1983) 405--414.
\newblock \href {http://dx.doi.org/10.1109/TNS.1983.4332300}
  {\path{doi:10.1109/TNS.1983.4332300}}.

\bibitem{Eremin2003}
V.~Eremin, J.~Bohm, S.~Roe, G.~Ruggiero, P.~Weilhammer, {T}he charge collection
  in single side silicon microstrip detectors, Nucl. Instr. $\&$ Meth. A 500
  (2003) 121--132.
\newblock \href {http://dx.doi.org/10.1016/S0168-9002(03)00330-9}
  {\path{doi:10.1016/S0168-9002(03)00330-9}}.

\bibitem{Verbit2005}
E.~Verbitskaya, et~al., {E}ffect of {SiO}$_2$ {P}assivating {L}ayer in
  {S}egmented {S}ilicon {P}lanar {D}etectors on the {D}etector {R}esponse, IEEE
  Trans. Nucl. Sci 52 (2005) 1877--1881.
\newblock \href {http://dx.doi.org/10.1109/TNS.2005.856907}
  {\path{doi:10.1109/TNS.2005.856907}}.

\bibitem{Poehlsen2013}
T.~Poehlsen, et~al., {C}harge losses in segmented silicon sensors at the
  {S}i-{SiO}$_2$ interface, Nucl. Instr. $\&$ Meth. A 700 (2013) 22--39.
\newblock \href {http://dx.doi.org/10.1016/j.nima.2012.10.063}
  {\path{doi:10.1016/j.nima.2012.10.063}}.

\bibitem{Poehlsen2013b}
T.~Poehlsen, et~al., {S}tudy of the accumulation layer and charge losses at the
  {S}i-{SiO}$_2$ interface in \textit{p}$^+$\textit{n}-silicon strip sensors,
  Nucl. Instr. $\&$ Meth. A 721 (2013) 26--34.
\newblock \href {http://dx.doi.org/10.1016/j.nima.2013.04.026}
  {\path{doi:10.1016/j.nima.2013.04.026}}.

\bibitem{Poehlsen2013c}
T.~Poehlsen, et~al., {T}ime dependence of charge losses at the {S}i-{SiO}$_2$
  interface in p$^+$n-silicon strip sensors, Nucl. Instr. $\&$ Meth. A 731
  (2013) 172--176.
\newblock \href {http://dx.doi.org/10.1016/j.nima.2013.03.035}
  {\path{doi:10.1016/j.nima.2013.03.035}}.

\bibitem{Ramo1939}
S.~Ramo, {C}urrents induced by electron motion, Proc. I.R.E. 27 (1939)
  584--585.
\newblock \href {http://dx.doi.org/10.1109/JRPROC.1939.228757}
  {\path{doi:10.1109/JRPROC.1939.228757}}.

\bibitem{Radeka1988}
V.~Radeka, {L}ow-{N}oise {T}echniques in {D}etectors, Annu. Rev. Nucl. Part.
  Sci. 38 (1988) 217--277.
\newblock \href {http://dx.doi.org/10.1146/annurev.ns.38.120188.001245}
  {\path{doi:10.1146/annurev.ns.38.120188.001245}}.

\bibitem{Lutz1999}
G.~Lutz, Semiconductor Radiation Detectors, Springer, Berlin, 1999.

\bibitem{He2001}
Z.~He, {R}eview of the {S}hockley-{R}amo theorem and its application in
  semiconductor gamma-ray detectors, Nucl. Instr. $\&$ Meth. A 463 (2001)
  250--267.
\newblock \href {http://dx.doi.org/10.1016/S0168-9002(01)00223-6}
  {\path{doi:10.1016/S0168-9002(01)00223-6}}.

\bibitem{Spieler2005}
H.~Spieler, Semiconductor Detector Systems, Oxford University Press, New York,
  NY, USA, 2005.

\bibitem{Peltola2014}
T.~Peltola, {C}harge collection efficiency simulations of irradiated silicon
  strip detectors, JINST 9 (2014) C12010.
\newblock \href {http://dx.doi.org/10.1088/1748-0221/9/12/C12010}
  {\path{doi:10.1088/1748-0221/9/12/C12010}}.

\bibitem{Bhardwaj2014}
A.~Bhardwaj, R.~Dalal, R.~Eber, T.~Eichhorn, K.~Lalwani, A.~Messineo,
  T.~Peltola, M.~Printz, K.~Ranjan, \textit{{S}imulation of {S}ilicon {D}evices
  for the {CMS P}hase {II T}racker {U}pgrade}, Compact Muon Solenoid, {CMS DN}
  -2014/016 (2015).

\bibitem{Peltola2015}
T.~Peltola, et~al., {A} method to simulate the observed surface properties of
  proton irradiated silicon strip sensors, JINST 10 (2015) C04025.
\newblock \href {http://dx.doi.org/10.1088/1748-0221/10/04/C04025}
  {\path{doi:10.1088/1748-0221/10/04/C04025}}.

\bibitem{Peltola2015r}
T.~Peltola,
  \href{pos.sissa.it/archive/conferences/254/031/VERTEX2015_031.pdf}{{Si}mulat%
ion of radiation-induced defects}, PoS 031 (2015) ({VERTEX2015}).
\newline\urlprefix\url{pos.sissa.it/archive/conferences/254/031/VERTEX2015_031%
.pdf}

\bibitem{Peltola2016p}
T.~Peltola, et~al., {C}haracterization of thin p-on-p radiation detectors with
  active edges, Nucl. Instr. $\&$ Meth. A 813 (2016) 139--146.
\newblock \href {http://dx.doi.org/10.1016/j.nima.2016.01.016}
  {\path{doi:10.1016/j.nima.2016.01.016}}.

\bibitem{Dalal2014}
R.~Dalal, et~al., \href{pos.sissa.it/}{{S}imulation of irradiated {S}i
  detectors}, PoS 030 (2014) ({VERTEX2014}).
\newline\urlprefix\url{pos.sissa.it/}

\bibitem{Eber2013}
R.~Eber,
  \href{http://ekp-invenio.physik.uni-karlsruhe.de/record/48328/files/EKP-2014%
-00012.pdf}{{I}nvestigations of new sensor designs and development of an
  effective radiation damage model for the simulation of highly irradiated
  silicon particle detectors}, Ph.D. thesis, Karlsruhe Institute of Technology
  (2013).
\newline\urlprefix\url{http://ekp-invenio.physik.uni-karlsruhe.de/record/48328%
/files/EKP-2014-00012.pdf}

\bibitem{Peltola2016}
T.~Peltola,
  \href{https://helda.helsinki.fi/bitstream/handle/10138/159441/numerica.pdf?s%
equence=1}{{N}umerical simulations of semiconductor radiation detectors for
  high-energy physics and spectroscopy applications}, Ph.D. thesis, University
  of Helsinki (2016).
\newline\urlprefix\url{https://helda.helsinki.fi/bitstream/handle/10138/159441%
/numerica.pdf?sequence=1}

\bibitem{Ruggiero2003}
G.~Ruggiero, {S}ignal {G}eneration in highly irradiated silicon microstrip
  detectors for the {ATLAS} experiment, Ph.D. thesis, Dept. Phys. Astr.,
  University of Glasgow, Glasgow, U.K. (2003).

\bibitem{Sze1981}
S.~M. Sze, Physics of Semiconductor Devices, 2nd Edition, John Wiley \& Sons,
  New Jersey, 1981.

\bibitem{Shockley1952}
W.~Shockley, W.~T. Read, Jr., {S}tatistics of the recombinations of holes and
  electrons, Phys. Rev. 87.
\newblock \href {http://dx.doi.org/10.1103/PhysRev.87.835}
  {\path{doi:10.1103/PhysRev.87.835}}.

\bibitem{Burhop1952}
E.~H.~S. Burhop, The Auger Effect and Other Radiationless Transitions,
  Cambridge Monographs on Physics, Cambridge, 1952.

\bibitem{Zhang2012}
J.~Zhang, et~al., \href{pos.sissa.it/}{{X}-ray induced radiation damage in
  segmented p$^+$n silicon sensors}, PoS 019 (2012) ({VERTEX2012}).
\newline\urlprefix\url{pos.sissa.it/}

\bibitem{Moscatelli2016}
F.~Moscatelli, et~al., {C}ombined bulk and surface radiation damage effects at
  very high fluences in silicon detectors: {M}easurements and {TCAD}
  simulations, IEEE Trans. Nucl. Sci 63 (2016) 2716--2723.
\newblock \href {http://dx.doi.org/10.1109/TNS.2016.2599560}
  {\path{doi:10.1109/TNS.2016.2599560}}.

\bibitem{Cavalleri1971}
G.~Cavalleri, E.~Gatti, G.~Fabri, V.~Svelto, {E}xtension of {R}amo's theorem as
  applied to induced charge in semiconductor detectors, Nucl. Instr. $\&$ Meth.
  A 92 (1971) 137--140.
\newblock \href {http://dx.doi.org/10.1016/0029-554X(71)90235-7}
  {\path{doi:10.1016/0029-554X(71)90235-7}}.

\bibitem{Gatti1982}
E.~Gatti, G.~Padovini, V.~Radeka, {S}ignal evaluation in multielectrode
  radiation detectors by means of a time dependent weighting vector, Nucl.
  Instr. $\&$ Meth. A 193 (1982) 651--653.
\newblock \href {http://dx.doi.org/10.1016/0029-554X(82)90265-8}
  {\path{doi:10.1016/0029-554X(82)90265-8}}.

\end{thebibliography}

\end{document}